\documentclass[12pt, a4paper]{article}
\usepackage{epstopdf}

\usepackage[utf8]{inputenc}
\usepackage{amsmath, amssymb, amsfonts, amsthm}
\usepackage{graphicx}
\usepackage{tikz}
\usepackage{titlesec} 
\usetikzlibrary{shadows.blur}
\usetikzlibrary{shadows}
\usepackage{seqsplit}
\usepackage{hyperref}
\usepackage{geometry}
\usepackage{setspace} 
\usepackage{algorithm}
\usepackage{titlesec} 
\usepackage{algpseudocode}
\usepackage{physics}
\usepackage{siunitx}
\usepackage{tikz}
\usetikzlibrary{positioning,shapes,arrows}
\usetikzlibrary{positioning, arrows.meta, calc, decorations.pathreplacing}
\AtBeginDocument{\RenewCommandCopy\qty\SI}
\AtBeginDocument{\RenewCommandCopy\qty\SI}
\usepackage[numbers, sort&compress]{natbib}
\usepackage{titling}
\usepackage{graphicx}
\usepackage{amsmath}
\usepackage{amssymb}
\usepackage{booktabs}
\usepackage{bm}
\usepackage{hyperref}
\AtBeginDocument{\RenewCommandCopy\qty\SI}
\pretitle{\begin{center}\large\bfseries}
\posttitle{\end{center}}

\preauthor{\begin{center}\normalsize}
\postauthor{\end{center}}

\predate{\begin{center}\small}
\postdate{\end{center}}
\usepackage{tikz}
\usetikzlibrary{shapes.geometric, arrows.meta, positioning, calc, backgrounds, fit}

\usepackage{booktabs} 
\usepackage{caption}
\usepackage{subcaption}

\title{Physics Constrained Neural Collision Operators for Variable Hard Sphere Surrogates and Ab Initio Angle Prediction in Direct Simulation Monte Carlo}

\author{
Ehsan Roohi$^{1}$\thanks{Corresponding author: roohie@umass.edu},
Ahmad Shoja-sani$^{2}$,
Stefan Stefanov$^{3,4}$\\[2pt]
$^{1}$Mechanical and Industrial Engineering, University of Massachusetts Amherst,
160 Governors Dr., Amherst, MA 01003, USA\\$^{2}$Department of Mechanical Engineering, Ferdowsi University of Mashhad,
Mashhad, 9177948974, Iran\\
$^{3}$Institute of Mechanics, Bulgarian Academy of Sciences, Acad. G. Bontchev Str., 1113 Sofia, Bulgaria\\
$^{4}$Centre of Excellence in Informatics and Information and Communication Technologies, Sofia, Bulgaria
}

\date{\today}

\AtBeginDocument{\RenewCommandCopy\qty\SI}

\begin{document}

\maketitle

\begin{abstract}
The Direct Simulation Monte Carlo (DSMC) method is the gold standard for non-equilibrium rarefied gas dynamics, yet its computational cost can be prohibitive, especially for near-continuum regimes and high-fidelity \emph{ab initio} potentials. This work develops a unified, physics-constrained neural-operator framework that accelerates DSMC while preserving physical invariants and stochasticity required for long-time kinetic simulations. 
First, we introduce a local neural collision kernel replacing the phenomenological Variable Hard Sphere (VHS) model. To overcome the variance suppression and artificial cooling inherent to purely deterministic regression surrogates, we augment inference with a physics-constrained stochastic layer. Controlled latent-noise injection restores thermal fluctuations, while cell-wise moment-matching strictly enforces momentum and kinetic-energy conservation. Remarkably, this operator exhibits zero-shot spatial and thermodynamic generalization: a model trained exclusively on 1D Couette flow accurately simulates a complex 2D lid-driven cavity, capturing high-order non-equilibrium moments without retraining.
Second, to bypass the extreme cost of quantum-mechanical scattering, we develop a dedicated \emph{ab initio} neural operator for the J\"ager interaction potential. Trained via a \emph{physics harvesting} strategy on large-scale collision pairs, it efficiently captures the high-energy scattering dynamics dominating hypersonic regimes. Validated on a Mach~10 rarefied argon flow over a cylinder, the framework reproduces transport behaviors and shock features with high fidelity, achieving an approximate 20\% cost reduction relative to direct numerical integration. 
Collectively, this work establishes physics-constrained neural operators as accurate, stable, and efficient drop-in surrogates for DSMC collision dynamics across both engineering VHS setups and \emph{ab initio} hypersonic simulations.
\end{abstract}

\subsection*{Keywords}

Direct Simulation Monte Carlo, Rarefied Gas Dynamics, Neural Collision Operator, ab initio potential, Local neural collision operator, hypersonic cylinder

\newpage

\section{Introduction}

The accurate simulation of gas flows in the rarefied regime remains a cornerstone challenge in modern aerospace and mechanical engineering, particularly for the design of re-entry vehicles, high-altitude aerodynamics, and micro/nano-electromechanical systems (MEMS/NEMS). In these regimes, characterized by a Knudsen number $Kn > 0.1$, the continuum hypothesis inherent to the Navier-Stokes-Fourier equations breaks down, necessitating a kinetic description governed by the Boltzmann equation \cite{vincenti1967introduction, bird1994molecular}. Historically, the Direct Simulation Monte Carlo (DSMC) method, pioneered by Bird \cite{bird2013dsmc}, has established itself as the gold standard for solving the Boltzmann equation stochastically. By tracking the motion and collisions of representative particles, DSMC accurately captures non-equilibrium effects that continuum solvers miss, as detailed in recent comprehensive reviews \cite{roohi2025advances}.

Despite its unrivaled physical fidelity, DSMC faces significant computational bottlenecks, particularly in the near-continuum regime where collision rates become exorbitant \cite{gallis2014direct, moore2019direct}. To address the trade-off between accuracy and cost, various strategies have been employed, ranging from hybrid methods coupling DSMC with Navier-Stokes solvers \cite{darbandi2013hybrid} to adaptive sub-cell strategies for shock tubes \cite{roohi2026time} and beyond \cite{balaj2014investigation}. While approaches like Direct Molecular Simulation (DMS) offer high fidelity by circumventing empirical collision kernels like the Variable Hard Sphere (VHS) model, they remain computationally prohibitive for engineering-scale problems. Consequently, the research community has pivoted towards Scientific Machine Learning (SciML) as a potent tool to approximate complex physical interactions with high efficiency \cite{karniadakis2021physics, cuomo2022scientific}.

Recent advancements have demonstrated the efficacy of deep learning in fluid dynamics, including methods for Reynolds-averaged Navier-Stokes simulations \cite{thuerey2020deep} and steady flow approximation using Convolutional Neural Networks \cite{guo2016convolutional}. In the realm of solving partial differential equations, Physics-Informed Neural Networks (PINNs) \cite{raissi2019physics, farea2024understanding} and Deep Operator Networks (DeepONets) \cite{lu2021deeponet, wang2021learning} have shown immense promise, with comprehensive reviews highlighting their applications in complex fluid dynamics \cite{zhao2024comprehensive}. Specific to rarefied gas dynamics, neural networks have been utilized to accelerate Boltzmann equation solutions \cite{xiao2021using, miller2022neural}, model BGK relaxation processes \cite{deflorio2021physics, deflorio2022physics}, and learn collision operators for lattice Boltzmann methods \cite{corbetta2023toward, liu2024physics}. Furthermore, PINNs have successfully tackled inverse problems such as solving the Vlasov-Poisson equation \cite{zhang2023physics}. More recent studies have expanded these efforts to include surrogate modeling for DSMC solutions \cite{roohi2026data, tatsios2025dsmc}, prediction of spectral responses \cite{chen2025prediction}, and the analysis of rarefied flows in complex geometries such as cylinder arrays \cite{tucny2025physics} and nanochannels \cite{to2025size}. Advanced operator learning techniques have also been applied to specific rarefied regimes, including micro-step flows \cite{roohi2025analysis} and micro-nozzles \cite{roohi2025shock}, utilizing physics-guided loss functions.

However, the direct replacement of a physical collision kernel with a neural network is non-trivial and presents fundamental challenges. A primary issue is the stochastic nature of gas collisions. Neural networks trained via standard Mean Squared Error (MSE) loss functions act as deterministic regressors. When applied to stochastic systems, such networks tend to predict the \textit{expectation value} of the output distribution, effectively filtering out the high-frequency variance essential to thermal fluctuations. In thermodynamic terms, this statistical averaging manifests as an artificial reduction in entropy, causing the system to undergo unphysical "cooling" or "freezing" over time. This phenomenon, known as ``Regression to the Mean,'' represents a critical barrier to the deployment of stable ML-driven kinetic solvers.

In response to these challenges, recent work by Ball et al.\cite{ball2025online} introduced an online optimization framework to calibrate collision models for DMS. While our work shares the overarching objective of integrating neural networks with kinetic solvers, the fundamental approach, network architecture, and the handling of physical constraints differ substantially. 
A crucial distinction in our methodology is the explicit restoration of thermodynamic consistency. Because scalar-prediction models do not directly manipulate velocity vectors, they do not encounter the ``Regression to the Mean'' phenomenon directly. In our fully neural approach, this issue is paramount. We address this by introducing a novel, deterministic \textit{Physics-Constrained Inference Layer} based on Moment Matching. This post-processing module explicitly enforces the conservation of momentum and energy \textit{after} the neural inference. This ensures thermodynamic consistency and prevents the unphysical energy drift that plagues pure ML surrogates, a rigorous constraint absent in scalar-prediction models. Furthermore, regarding stochasticity, we treat molecular collisions as a latent variable problem. We employ an \textit{Input-Noise Injection} strategy, training the network to map a combined vector of deterministic velocities and random noise to a valid post-collision state. This allows the network to internalize the stochastic mapping and effectively learn the conditional probability distribution of collision outcomes end-to-end. Finally, to mitigate generalization errors without the computational overhead of online retraining used by Ball et al. \cite{ball2025online}, we propose a robust offline strategy termed \textit{Physics Harvesting}. By curating a high-fidelity training dataset from the Simplified Bernoulli Trial (SBT) kernel that specifically targets the high-energy tails of the distribution—characteristic of strong shock waves—we demonstrate that a single offline-trained model can generalize accurately in the hypersonic regimes, e.g., Mach 10.

In summary, we develop a unified physics-constrained neural-operator framework that accelerates DSMC by learning the microscopic collision law while preserving the robust stochastic structure of the classical algorithm. For phenomenological collisions, we replace the VHS scattering kernel with a neural collision operator and demonstrate geometry-independent generalization: a model trained only on 1D Couette-flow collision statistics is deployed, without any re-training, to a 2D lid-driven cavity and accurately reproduces not only the primary fields (velocity, temperature, density, and pressure) but also higher-order non-equilibrium moments such as shear stress and heat flux. To prevent the well-known ``regression-to-the-mean'' pathology of deterministic surrogates, we introduce a thermodynamically consistent post-processing/inference layer that restores fluctuations and enforces exact conservation constraints (cell-wise moment matching and/or collision-wise energy-shell projection), thereby eliminating spurious cooling/heating and enabling long-time stable simulations. Building on this physics-aware surrogate paradigm, we address the prohibitive cost of quantum-mechanical scattering by constructing a compact MLP surrogate for the J\"ager \textit{ab initio} Argon--Argon potential using a targeted physics-harvesting dataset spanning extreme collision energies. We then embed this model into a high-performance DSMC implementation via a GPU-resident neural table look-up strategy, reducing costly on-the-fly integral evaluations to constant-time memory fetches. The resulting solver is rigorously validated on rarefied hypersonic Argon flow over a cylinder at Mach~10, where it reproduces shock stand-off distance, post-shock relaxation, and surface quantities (e.g., skin-friction) with fidelity comparable to exact \textit{ab initio} integration, while delivering substantial computational speedups. Collectively, these contributions show that physics-constrained neural operators can serve as accurate, stable, and efficient drop-in surrogates for DSMC collision dynamics across both geometry-generalizing VHS settings and high-fidelity \textit{ab initio} hypersonic regimes.

\section{Methodology}

\subsection{DSMC collision step and scope of replacement}
DSMC advances a rarefied gas by alternating
(i) a free-flight (streaming) step, in which representative particles move ballistically
under the imposed boundary conditions, and (ii) a collision step, in which
binary interactions are sampled stochastically so as to reproduce the Boltzmann
collision integral in an expected (Monte Carlo) sense~\cite{bird1994molecular}.
In the standard Bird framework, the collision stage is performed locally by
partitioning the domain into collision cells whose linear size is of the order of,
or smaller than, the local mean free path. Within each cell, candidate pairs are
drawn and accepted according to a prescribed collision-rate algorithm (e.g.,
No-Time-Counter, NTC), ensuring the correct rate scaling
$\nu \propto n\,\sigma\,v_r$ and therefore the correct mean free path and transport behavior.

The present work follows this classical DSMC structure and is intentionally
\emph{surgical}: we preserve the robust cell-based pair selection and acceptance logic
(e.g., NTC/SBT-style rate control), as well as the standard center-of-mass velocity update,
and replace only the phenomenological scattering rule that maps a pre-collision
relative velocity $\mathbf{v}_r$ to a post-collision update $\mathbf{v}'_r$.
This design choice isolates the dominant modeling/computational bottleneck---the
microscopic scattering law---while keeping the proven DSMC machinery for locality,
rates, and sampling unchanged. The next subsection introduces our cell-local neural
collision operator, which is invoked only for accepted collision pairs inside each
collision cell and thus remains fully compatible with the DSMC collision partition.

\subsection{Cell-Based Neural Collision Operator (Conditional MLP)}
\label{sec:neural_collision}

We replace the stochastic angular sampling of the DSMC collision operator with a lightweight
\emph{conditional neural surrogate} that predicts the post-collision relative velocity direction for each accepted collision pair.
Importantly, the surrogate is \emph{pairwise} and is invoked only for collision candidates selected within each DSMC collision cell,
thereby preserving the standard locality assumptions of DSMC.

\paragraph{Cell partition and collision-pair selection.}
The physical domain is partitioned into $N_c$ collision cells $\{C_k\}_{k=1}^{N_c}$.
Let cell $C_k$ contain $N_k$ particles with index set $\mathcal{I}_k$.
We follow the standard No-Time-Counter (NTC) style selection within each cell:
a number of candidate pairs is drawn proportional to $N_k(N_k-1)$ and a cell-wise upper bound
$(\sigma v_r)_{\max,k}$ of the collision rate.
For each trial, two particles $(i,j)\in\mathcal{I}_k$ are drawn uniformly and accepted with probability proportional to
$\sigma\,v_r$, where $v_r=\|\mathbf{v}_i-\mathbf{v}_j\|_2$ is the relative speed and $\sigma$ is the (here constant) reference cross section.

\paragraph{Neural surrogate input/output.}
For an accepted pair $(i,j)$ we define the relative velocity vector
\begin{equation}
\mathbf{v}_r = \mathbf{v}_i-\mathbf{v}_j \in \mathbb{R}^3,
\qquad
v_r = \|\mathbf{v}_r\|_2.
\end{equation}
To model the intrinsic stochasticity of molecular scattering while keeping inference deterministic,
we provide the network with an explicit random latent vector
$\boldsymbol{\xi}\in\mathbb{R}^3$ sampled independently for each collision event:
\begin{equation}
\boldsymbol{\xi}\sim\mathcal{N}(\mathbf{0},\mathbf{I}_3).
\end{equation}
We normalize the relative velocity by a global scale $\sigma_v>0$ (precomputed from training data)
and form the 6D network input
\begin{equation}
\mathbf{x}=
\left[
\frac{\mathbf{v}_r}{\sigma_v}
\;\; \oplus \;\;
\boldsymbol{\xi}
\right]\in\mathbb{R}^6,
\label{eq:mlp_input}
\end{equation}
where $\oplus$ denotes concatenation.
The neural model is a multilayer perceptron (MLP) $f_\theta:\mathbb{R}^6\to\mathbb{R}^3$ that predicts an \emph{unnormalized}
post-collision relative-velocity vector
\begin{equation}
\widehat{\mathbf{v}}_r' = f_\theta(\mathbf{x})\in\mathbb{R}^3.
\label{eq:mlp_output}
\end{equation}
In our implementation, $f_\theta$ consists of fully connected layers with ReLU activations.

\subsection{Hard Conservation via Projection and Pair Update}
\label{sec:projection_update}

\paragraph{Energy-shell projection for exact pairwise energy conservation.}
Small inference errors in the magnitude $\|\widehat{\mathbf{v}}_r'\|_2$ can accumulate over many collisions.
We therefore enforce exact pairwise kinetic-energy conservation in the relative frame by projecting the network output
onto the \emph{energy shell} defined by the pre-collision relative speed $v_r$:
\begin{equation}
\mathbf{u} = \frac{\widehat{\mathbf{v}}_r'}{\|\widehat{\mathbf{v}}_r'\|_2+\varepsilon},
\qquad
\mathbf{v}_r' = v_r\,\mathbf{u},
\label{eq:energy_shell_projection}
\end{equation}
where $\varepsilon>0$ is a small safeguard.

\paragraph{Velocity update in the center-of-mass frame.}
Let the center-of-mass velocity be
\begin{equation}
\mathbf{v}_{cm}=\frac{1}{2}\left(\mathbf{v}_i+\mathbf{v}_j\right).
\end{equation}
The post-collision velocities are updated by
\begin{align}
\mathbf{v}_i^{new} &= \mathbf{v}_{cm} + \frac{1}{2}\mathbf{v}_r',\\
\mathbf{v}_j^{new} &= \mathbf{v}_{cm} - \frac{1}{2}\mathbf{v}_r'.
\label{eq:v_update_pair}
\end{align}
This guarantees conservation of pair momentum and (through \eqref{eq:energy_shell_projection}) conservation of pair kinetic energy.

\subsection{Thermodynamic Stability: Why Conditional Noise is Needed}
A fundamental challenge in replacing stochastic physical kernels with neural networks is the "Regression to the Mean" phenomenon. Standard regressors trained with $L_2$ loss functions tend to approximate the conditional expectation of the target distribution, $E[\mathbf{v}_r' | \mathbf{v}_r]$, which effectively acts as a low-pass filter on thermal fluctuations. In kinetic simulations, this variance suppression manifests as an unphysical entropy loss, leading to "cooling" or freezing of the velocity distribution over time.

To restore the stochastic nature of the collision integral, we formulate the scattering process as a conditional generative task. The neural collision operator is defined by the mapping:
\begin{equation}
    \mathbf{\hat{v}}_r' = f_\theta(\mathbf{v}_r, \xi), \quad \xi \sim \mathcal{N}(0, \mathbf{I}_3)
\end{equation}
where $\xi$ is a latent random variable. This architecture allows the network to learn a conditional probability density function $P(\mathbf{v}_r' | \mathbf{v}_r, \xi)$ rather than a deterministic point estimate. 

The choice of a Gaussian distribution for the latent noise $\xi$ is physically motivated by the following:
Isotropy and Symmetry: The Gaussian distribution is rotationally invariant, which aligns with the expected isotropic scattering behavior in the center-of-mass frame for the VHS model.
Maxwell-Boltzmann Consistency: Since the components of velocity in a gas at equilibrium follow a normal distribution (Maxwellian), providing a Gaussian latent space $\xi \in \mathbb{R}^3$ ensures that the network has a sufficiently rich and compatible basis to reconstruct the high-frequency fluctuations characteristic of the Maxwell-Boltzmann distribution.
Stability: The use of $\xi \sim \mathcal{N}(0, \mathbf{I})$ prevents the collapse of the post-collision velocity variance, ensuring that the simulated gas maintains its target temperature and avoids the unphysical energy drift typically observed in purely deterministic ML surrogates.

\subsection{Optional Cell-Wise Energy Drift Correction}
\label{sec:cell_energy_clamp}

\begin{figure}[htbp]
\centering
\resizebox{0.92\textwidth}{!}{%
\begin{tikzpicture}[
    node distance=1.2cm and 1.6cm,
    box/.style ={rectangle, draw=black!70, fill=white, rounded corners, thick,
                 minimum width=4.2cm, minimum height=1.1cm, align=center, font=\sffamily\small},
    sbox/.style ={rectangle, draw=black!50, fill=gray!5, rounded corners, thick,
                 minimum width=4.2cm, minimum height=0.95cm, align=center, font=\sffamily\small},
    arrow/.style ={thick, ->, >=stealth, color=black!80, line width=1.2pt},
    note/.style ={font=\footnotesize, align=left, color=black!70}
]

\node[box] (cell) {Collision cell $C_k$\\select candidate pairs $(i,j)$ (NTC-style)};
\node[box, below=of cell] (feat) {Build features\\$\mathbf{v}_r=\mathbf{v}_i-\mathbf{v}_j$, $v_r=\|\mathbf{v}_r\|_2$\\$\boldsymbol{\xi}\sim\mathcal{N}(\mathbf{0},\mathbf{I})$\\$\mathbf{x}=[\mathbf{v}_r/\sigma_v \;\oplus\; \boldsymbol{\xi}]$};
\node[box, below=of feat] (mlp) {Conditional MLP surrogate\\$\widehat{\mathbf{v}}_r' = f_\theta(\mathbf{x})$};
\node[sbox, below=of mlp] (proj) {Hard projection (energy shell)\\$\mathbf{v}_r' = v_r\,\dfrac{\widehat{\mathbf{v}}_r'}{\|\widehat{\mathbf{v}}_r'\|_2+\varepsilon}$};
\node[box, below=of proj] (update) {Pair update\\$\mathbf{v}_{cm}=\tfrac12(\mathbf{v}_i+\mathbf{v}_j)$\\$\mathbf{v}_i^{new}=\mathbf{v}_{cm}+\tfrac12\mathbf{v}_r'$\\$\mathbf{v}_j^{new}=\mathbf{v}_{cm}-\tfrac12\mathbf{v}_r'$};
\node[sbox, below=of update] (clamp) {Optional cell-wise energy clamp\\$\alpha_k=\sqrt{E_k^{pre}/E_k^{post}}$, $\mathbf{v}\leftarrow \alpha_k\mathbf{v}$};

\draw[arrow] (cell) -- (feat);
\draw[arrow] (feat) -- (mlp);
\draw[arrow] (mlp) -- (proj);
\draw[arrow] (proj) -- (update);
\draw[arrow] (update) -- (clamp);

\node[note, right=0.4cm of mlp] {Noise is an \emph{input} latent\\(prevents variance collapse).};
\node[note, right=0.4cm of proj] {Enforces $|\mathbf{v}_r'|=|\mathbf{v}_r|$\\exactly.};

\end{tikzpicture}
}
\caption{Schematic of the neural collision operator. Collisions are selected locally in each DSMC cell. For each accepted pair, a conditional MLP predicts the post-collision relative velocity direction given the normalized pre-collision relative velocity and an explicit random latent vector. A hard projection enforces exact pairwise energy conservation before updating particle velocities.}
\label{fig:neural_collision_schematic}
\end{figure}
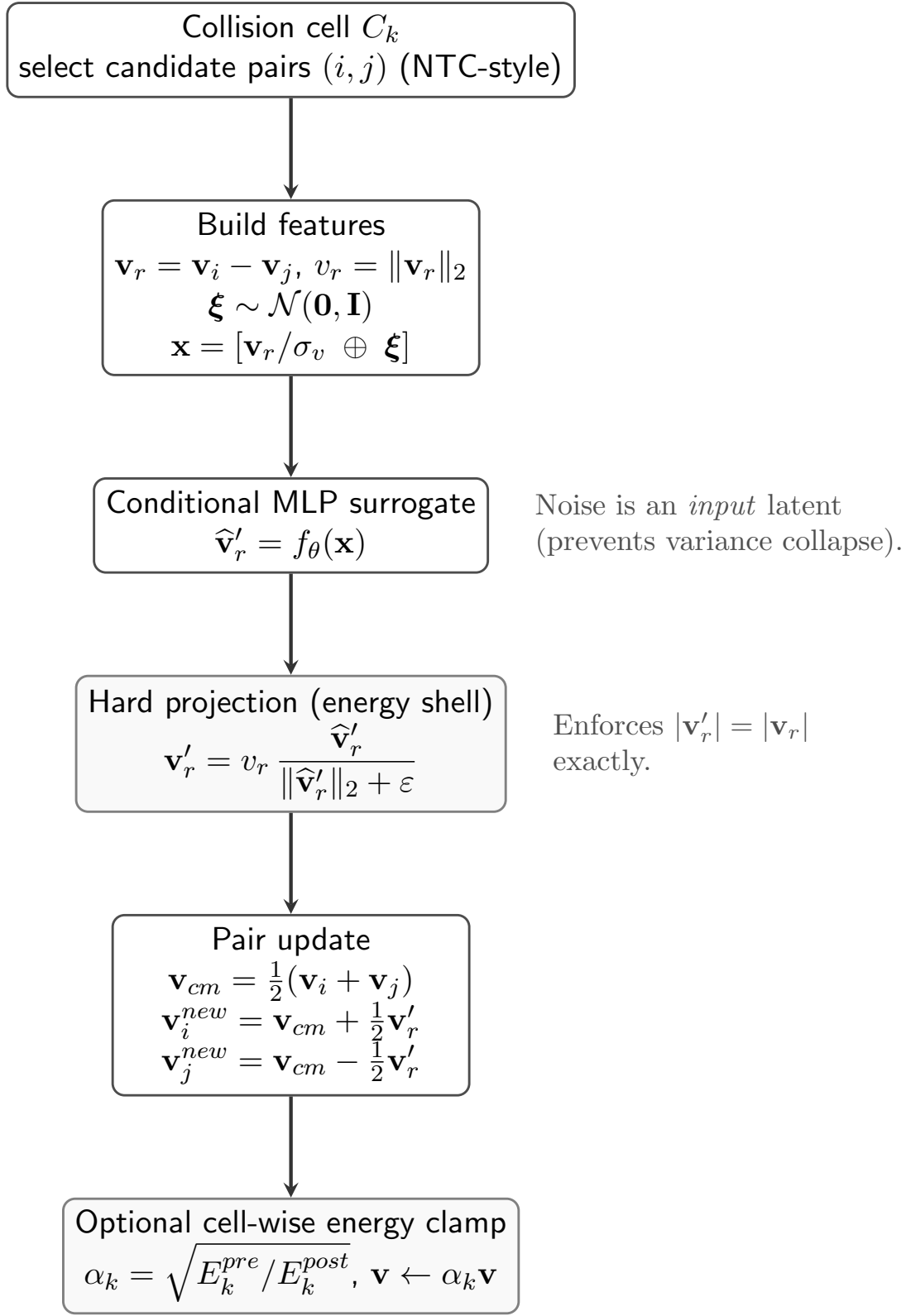

Although each collision enforces exact pairwise invariants, mild drift can still appear at the cell level due to finite sampling,
random acceptance, and floating-point round-off in long runs. We optionally apply a lightweight cell-wise energy correction.
For each cell $C_k$, let
\begin{equation}
E_k^{pre}=\sum_{i\in\mathcal{I}_k}\|\mathbf{v}_i\|_2^2,
\qquad
E_k^{post}=\sum_{i\in\mathcal{I}_k}\|\mathbf{v}_i^{new}\|_2^2,
\end{equation}
computed immediately before and after processing all collisions in the cell.
If $E_k^{post}>0$, we rescale all velocities in the cell by a single factor
\begin{equation}
\alpha_k=\sqrt{\frac{E_k^{pre}}{E_k^{post}}},
\qquad
\mathbf{v}_i \leftarrow \alpha_k\,\mathbf{v}_i,
\quad i\in\mathcal{I}_k,
\label{eq:cell_energy_clamp}
\end{equation}
which clamps the cell energy back to $E_k^{pre}$ up to round-off. The schematic of the developed machine learning algorithm is shown in Fig.~\ref{fig:neural_collision_schematic}. 

\subsection{Computational Framework}
\label{sec:comput_opt}

The surrogate is invoked only for collision candidates selected within each collision cell. 

\paragraph{(1) Spatial decomposition and local processing.}
Particles are sorted by cell index so that particles in the same cell are contiguous in memory, enabling linear scans over each cell.

\paragraph{(2) Batch inference per cell.}
For each cell, we assemble a batch of accepted collision pairs and evaluate $f_\theta$ in a vectorized manner on the batch,
reducing overhead and improving cache locality.

\paragraph{(3) Parallel execution over cells.}
Cell-wise collision processing is embarrassingly parallel and can be distributed across CPU threads (or GPU blocks in a CUDA implementation).

For training the network, approximately $200,000$ collision samples were harvested. This dataset represents the true statistical distribution of relative velocities in a non-equilibrium flow, including high-energy tail collisions that drive viscous heating in shear driven flows considered in this study.

The network architecture consists of a Multi-Layer Perceptron (MLP) with an input layer of size 6 (3 normalized velocity components + 3 noise components) and three hidden layers of width 256 with ReLU activation. The model was trained to minimize the Mean Squared Error (MSE) between the predicted and ground-truth scattering directions. By decoupling the magnitude $|\mathbf{v}_r|$ from the inference, the model achieves \textit{scale invariance}, ensuring robustness across wide temperature ranges.

\section{Results and Discussion}
\subsection{Relaxation to Equilibrium}
The simulation of flow relaxation to equilibrium was conducted for Argon gas ($m = 6.63 \times 10^{-26}$ kg) with an initial target temperature of $T_{init}=273.0$ K. 
The most critical validation for any thermodynamic simulation is its ability to maintain equilibrium. 
The final equilibrium temperature achieved by the solver was $274.41$ K, which is quite close to the target temperature of $273.0$ K. This indicates an absolute deviation of only $1.41$ K, confirming the high fidelity of the energy conservation mechanism.

\begin{figure}[h!]
    \centering
    \includegraphics[width=0.9\textwidth]{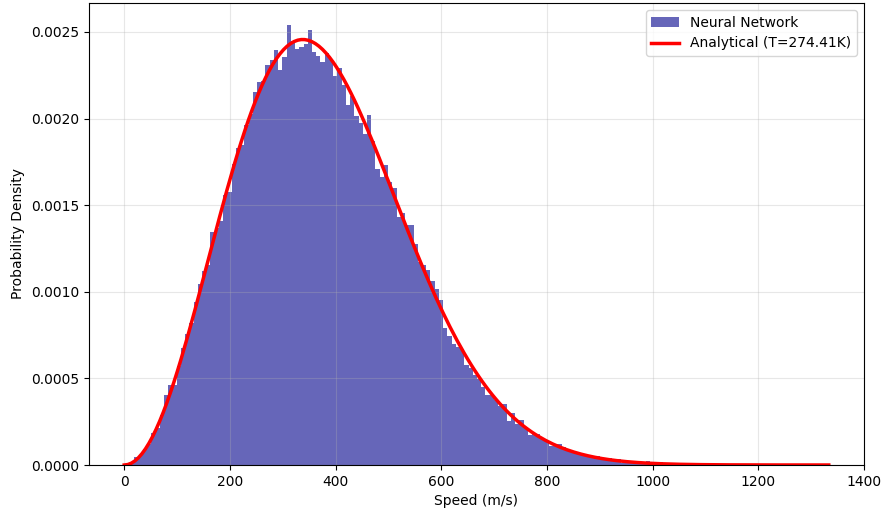}
    \caption{Comparison of the particle speed distribution from the Hybrid ML-DSMC solver (Blue Histogram) versus the analytical Maxwell-Boltzmann distribution at $T=274.41$ K (Red Line). The histogram exhibits an excellent fit with the theoretical curve, validating the successful restoration of the velocity distribution's shape and width through noise injection and energy rescaling.}
    \label{fig:maxwell}
\end{figure}

The core objective was to accurately reproduce the Maxwell-Boltzmann speed distribution. Figure \ref{fig:maxwell} illustrates the comparison between the simulated particle speeds (histogram) and the analytical Maxwell-Boltzmann distribution at the final equilibrium temperature.

As evident from Figure \ref{fig:maxwell}, the histogram of simulated particle speeds suitably overlays the theoretical Maxwell-Boltzmann curve. This precise agreement confirms that the combination of noise injection (to restore variance) and energy rescaling (to constrain the distribution's width) has successfully mitigated the "Regression to the Mean" phenomenon, yielding a thermodynamically consistent velocity distribution. 

\begin{figure}[h!]
    \centering
    \includegraphics[width=0.99\textwidth]{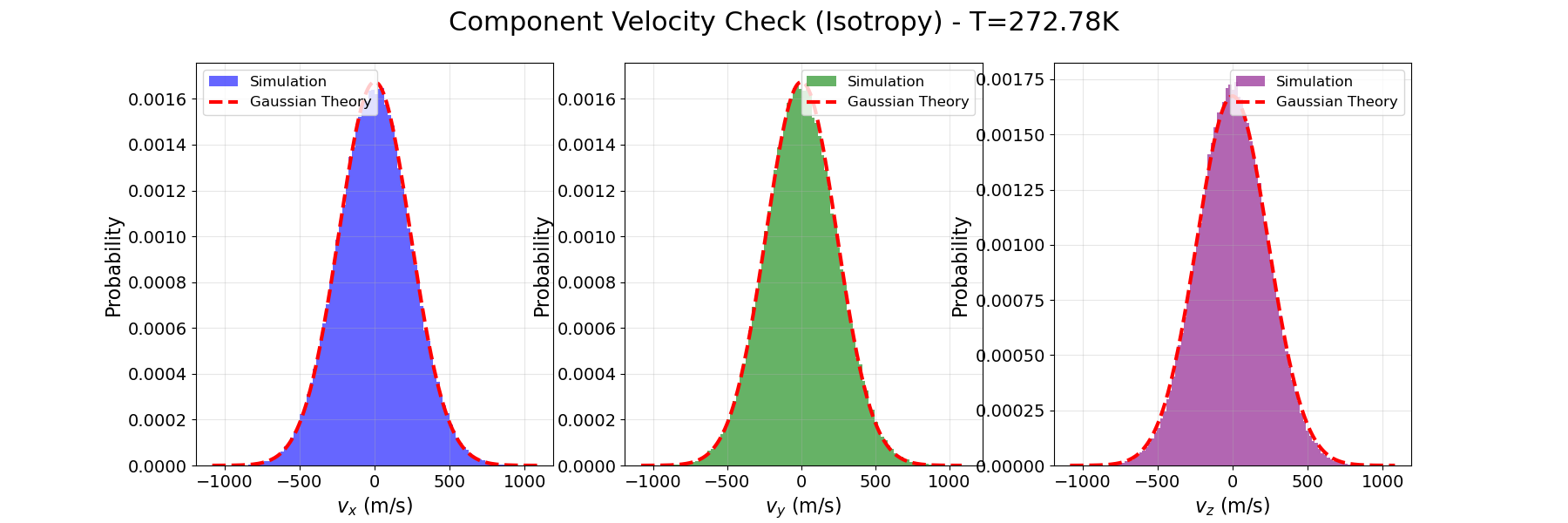}
    \caption{Distribution of individual velocity components ($v_x, v_y, v_z$) obtained from the simulation (histograms) compared against the theoretical Gaussian distribution (red dashed line) at $T=272.78$ K. The strong agreement for all three components confirms the isotropic nature of the simulated gas and the correct treatment of velocity directions by the DNN.}
    \label{fig:components}
\end{figure}

Beyond the overall speed distribution, it is crucial to verify that the gas's properties are isotropic—meaning uniform in all spatial directions. This is validated by examining the distribution of individual velocity components ($v_x, v_y, v_z$). Each component should follow a Gaussian (normal) distribution with a zero mean. Figure \ref{fig:components} presents these distributions.

Figure \ref{fig:components} clearly demonstrates that the simulated distributions for $v_x$, $v_y$, and $v_z$ align  suitably with their respective theoretical Gaussian curves. This verifies the isotropic behavior of the gas and indicates that the DNN, combined with the centering and energy rescaling steps, correctly models the directional aspects of particle velocities, preventing any artificial bias in specific spatial dimensions.

A quantitative assessment of the statistical moments further reinforces the accuracy of the solver. We compared the mean speed from the simulation with its theoretical value at $T=272.78$ K. The theoretical mean speed for a Maxwell-Boltzmann distribution is given by:
\begin{equation}
    \langle v \rangle_{theoretical} = \sqrt{\frac{8 k_B T}{\pi m}}
\end{equation}
Our statistical check yielded the following results:
Mean Speed (Simulation): $379.7416$ m/s,
Mean Speed (Theoretical): $380.2270$ m/s,
Relative Error: $0.1277\%$.
This extremely low error confirms that not only the shape but also the key statistical parameters of the velocity distribution are accurately reproduced by the hybrid DNN-DSMC solver. This level of precision is crucial for reliable predictions in kinetic gas theory.

\subsection{Rarefied Couette Flow}
To ensure the neural network accurately captures the collision dynamics of high-speed rarefied flows, we employed a "Data Harvesting" strategy rather than relying on synthetic, uniform distributions. A standard DSMC simulation of Couette flow of Argon gas ($Kn=0.05$, $U_{wall}=\pm 500$ m/s) was executed for a warm-up period of 500 time steps to allow the development of the macroscopic shear layer. Subsequently, we extracted the relative velocity vectors of particle pairs $(\mathbf{v}_r)$ that underwent successful collisions according to the No Time Counter (NTC) scheme. 
The trained neural network (NN) was integrated into the DSMC solver, replacing the phenomenological hard-sphere scattering kernel. To ensure long-term stability over $15,000$ time steps, a cell-level energy clamp was applied to prevent floating-point arithmetic drift, enforcing strict energy conservation $\sum v^2_{post} \equiv \sum v^2_{pre}$ at each time step.

\begin{figure}[h!]
    \centering
    \includegraphics[width=1.0\textwidth]{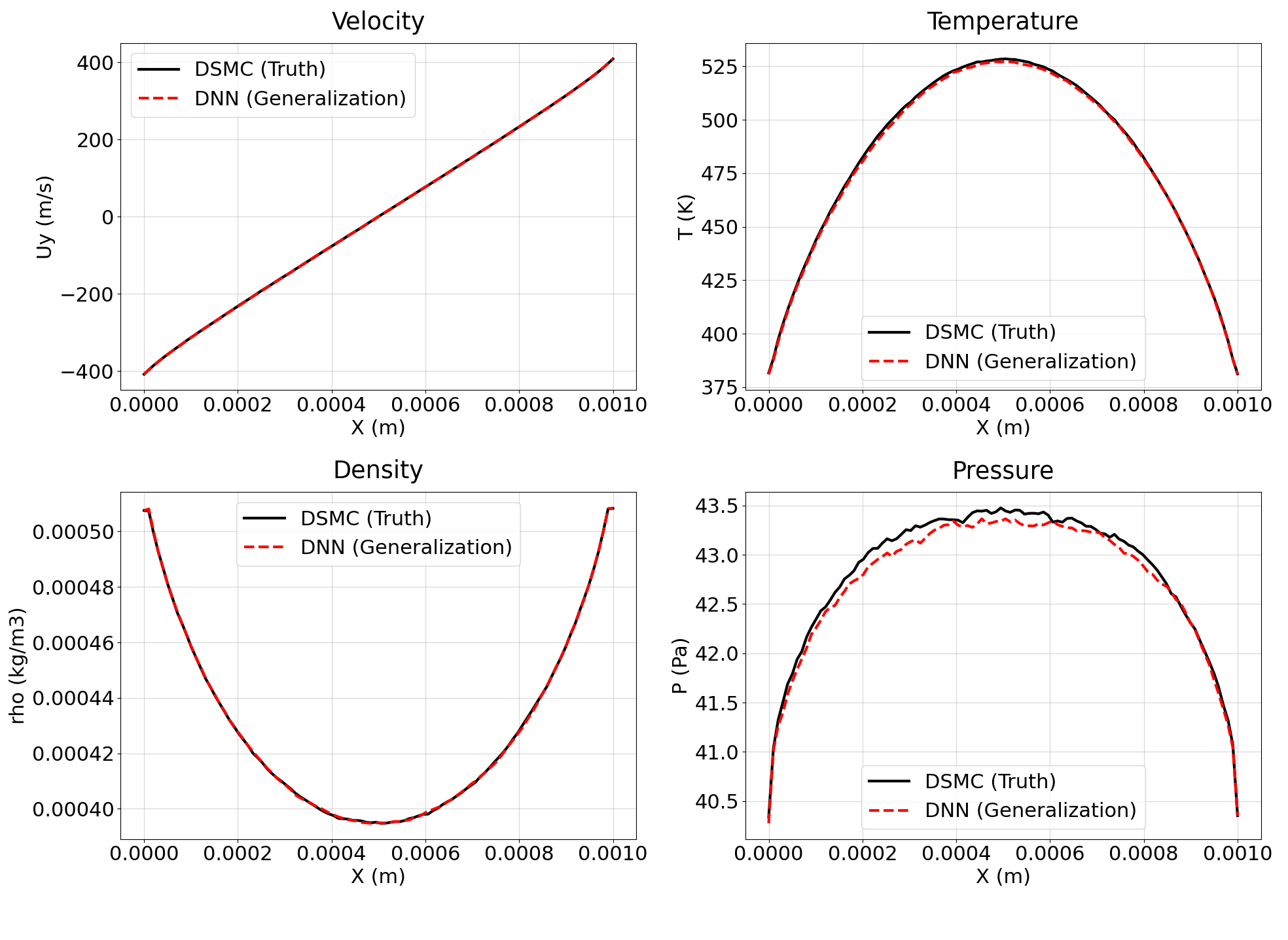}
    \caption{Macroscopic profiles for high-speed Couette flow ($U_w = 500$ m/s, $Kn=0.2$) comparing the Standard DSMC (Black Line) and the Hybrid DNN-DSMC (Red Dashed Line).}
    \label{fig:results}
\end{figure}

\begin{figure}[h!]
    \centering
    \includegraphics[width=1.0\textwidth]{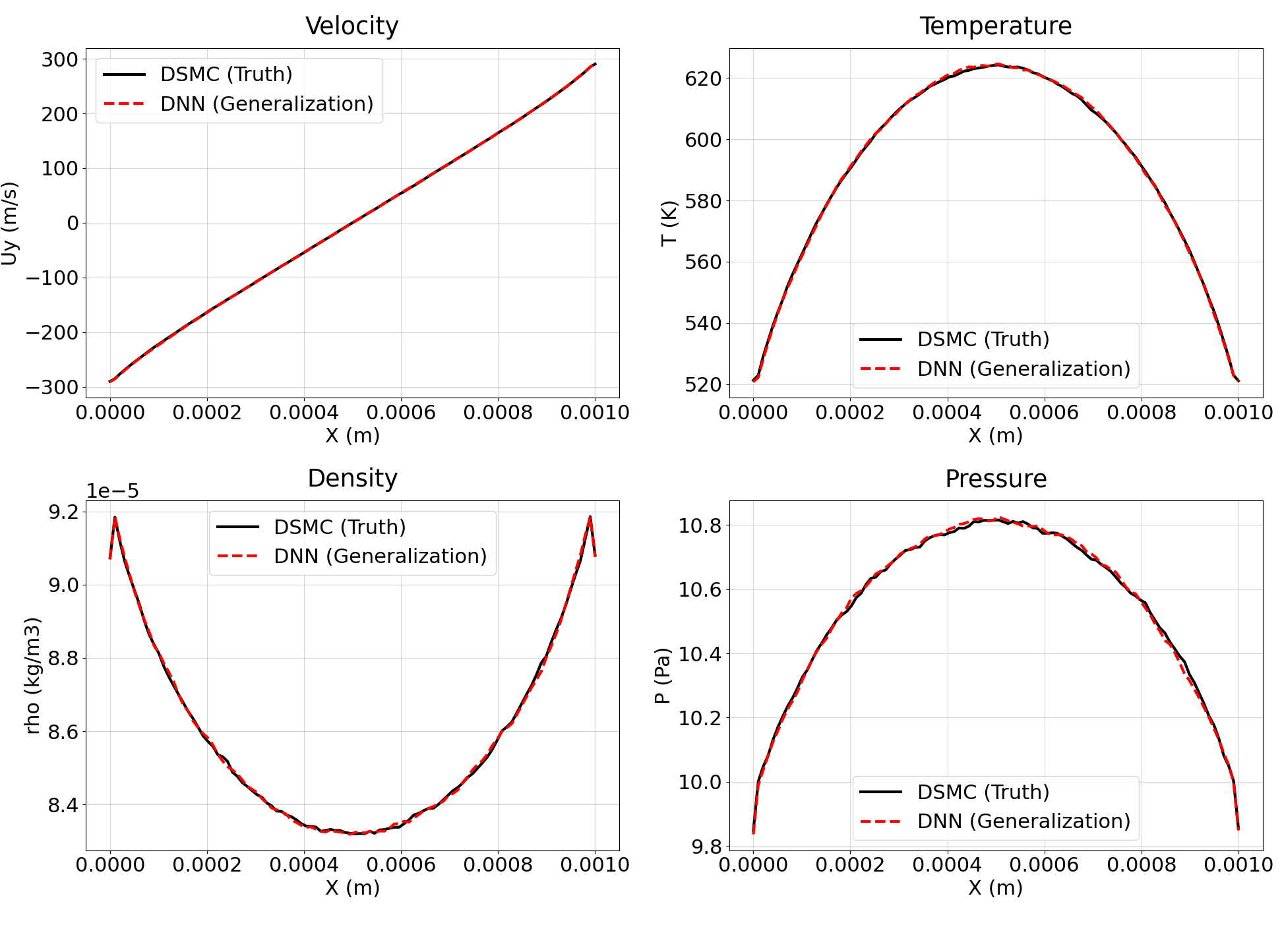}
    \caption{Validation of the DNN surrogate for 1D Couette flow in the transition regime. The simulation corresponds to a Knudsen number of $Kn=1.0$ and a wall velocity of $U_w = 500$ m/s.}
    \label{fig:couette_kn1}
\end{figure}

Figures \ref{fig:results} and \ref{fig:couette_kn1} present the macroscopic profiles obtained from the hybrid DNN-DSMC solver compared against the standard DSMC benchmark for the rarefied Couette flow at Kn=0.2 and Kn=1. The results demonstrate good agreement between the AI-driven solver and classical theory:
The velocity profile is linear in the bulk but exhibits distinct velocity slip at the walls ($|u_{gas}| < 500$ m/s). The DNN accurately reproduced the momentum transfer coefficient at the boundary.
The temperature profile follows a parabolic shape, peaking at $T \approx 525$ K. This confirms that the DNN has correctly learned the effective viscosity of the gas, as the heating rate is proportional to viscous dissipation.
Temperature, density and pressure profiles also show a good agreement between DSMC and DNN predictions.

\subsection{Generalization Test: Lid-Driven Cavity Flow}

To rigorously evaluate the predictive robustness of the physics-constrained neural operator, we perform a \textbf{zero-shot generalization test}. The conditional MLP, which was trained solely on 1D Couette flow data, is directly integrated into a 2D lid-driven cavity solver. This test is designed to verify whether the surrogate has internalized the fundamental, geometry-independent local collision laws or merely memorized the global flow field of the training case. Achieving accurate results in this configuration, characterized by multi-dimensional gradients and vortex structures, demonstrates a high degree of zero-shot spatial and thermodynamic generalization.

The simulation considers a square cavity of side length $L = 1.0$ mm filled with Argon gas. The top wall moves with a velocity of $U_{lid} = 100$ m/s, while the other three walls are stationary and diffuse at a temperature of $T_w = 273.15$ K. The Knudsen number is set to $Kn=0.1$, placing the flow in the slip/transition regime where non-equilibrium effects are significant.

To ensure high-fidelity statistical sampling and minimize thermal noise in this low-speed regime, strict numerical parameters were adopted for both the benchmark DSMC and the DNN-DSMC solver:
A uniform grid of $50 \times 50$ cells ($N_{cell}=2500$).
A total of $N_p = 250,000$ particles, ensuring approximately 100 particles per cell to resolve local moments accurately.
The simulation was run for $N_{steps} = 200,000$ time steps, with sampling starting after 10,000 steps to guarantee a statistically stationary state.
$\Delta t$ was chosen to be a fraction of the mean collision time to satisfy the DSMC stability criteria.

\begin{figure*}[htbp]
    \centering
    \includegraphics[width=0.9\textwidth]{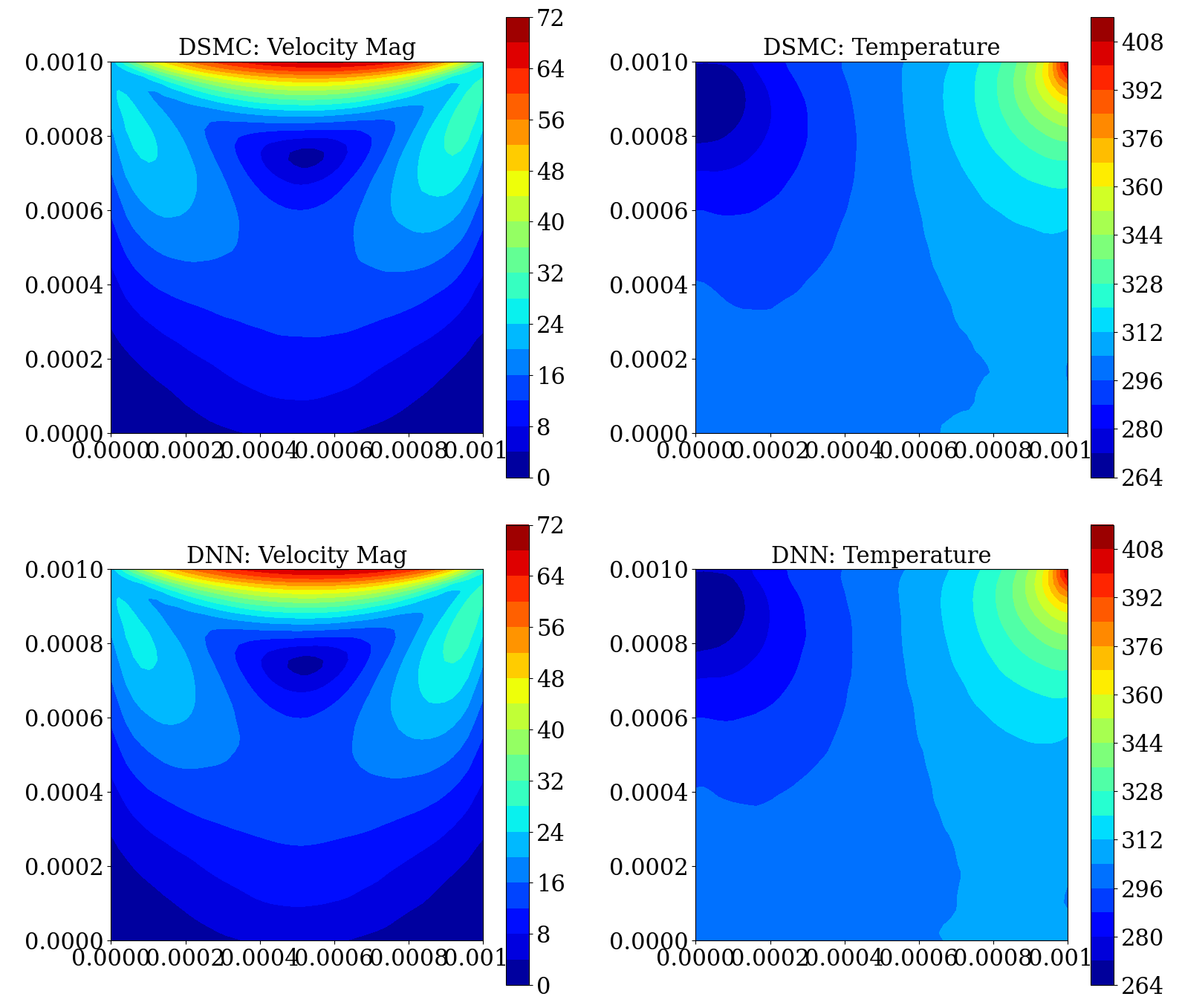}
    \caption{Comparison of macroscopic primary variables for 2D Lid-Driven Cavity flow ($Kn=0.1, U_{lid}=100$ m/s).}
    \label{fig:cavity_primary}
\end{figure*}

The macroscopic flow fields predicted by the DNNsurrogate are compared against the ground-truth DSMC results in Figure \ref{fig:cavity_primary}. The DNNsuccessfully reproduces the complex 2D flow features, including the primary vortex center, the steep velocity gradients near the moving lid, and the thermal gradients induced by viscous dissipation.

\begin{figure*}[htbp]
    \centering
    \includegraphics[width=0.9\textwidth]{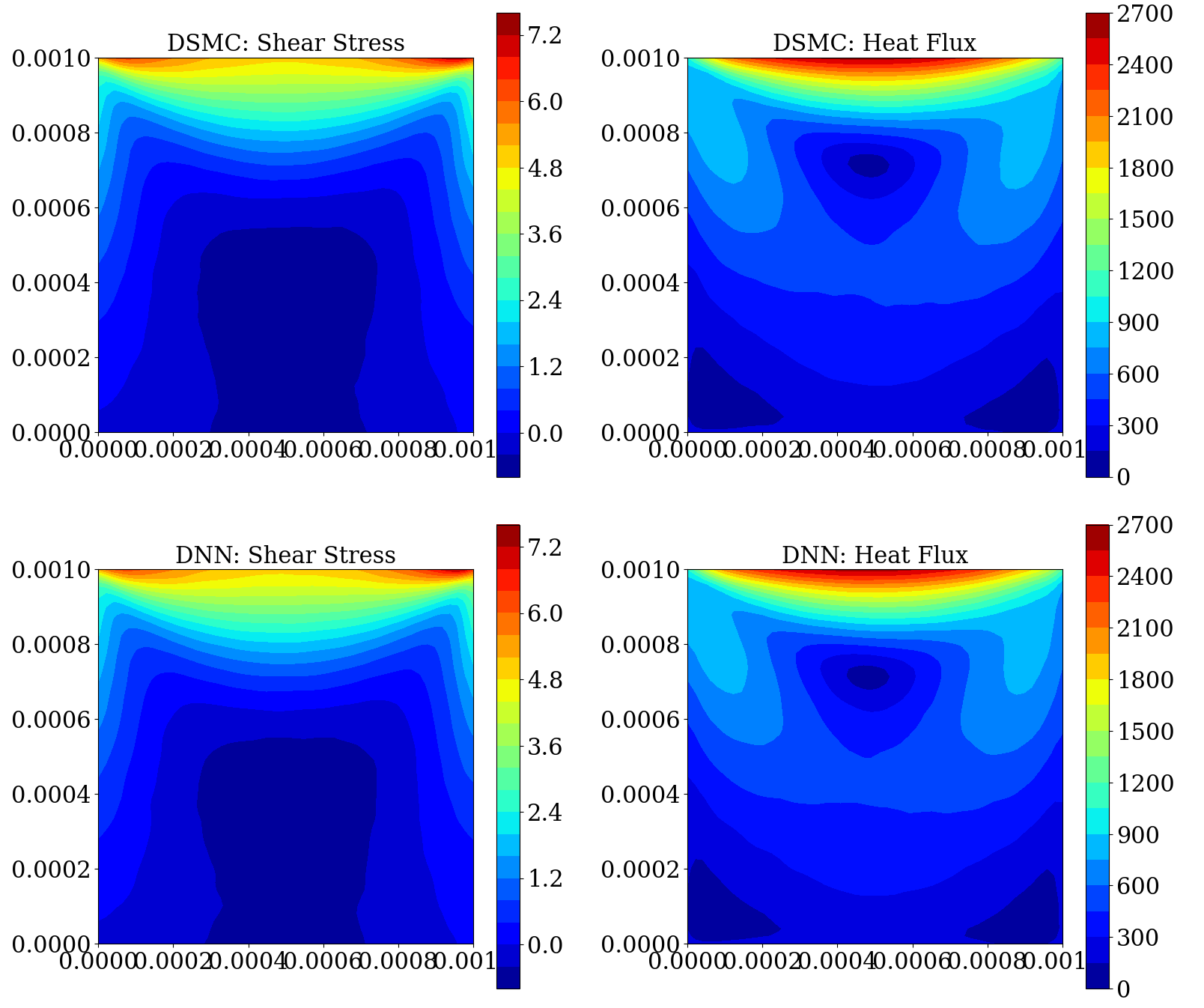}
    \caption{Comparison of higher-order non-equilibrium moments.}
    \label{fig:cavity_higher_order}
\end{figure*}

\begin{figure*}[htbp]
    \centering
    \includegraphics[width=1.0\textwidth]{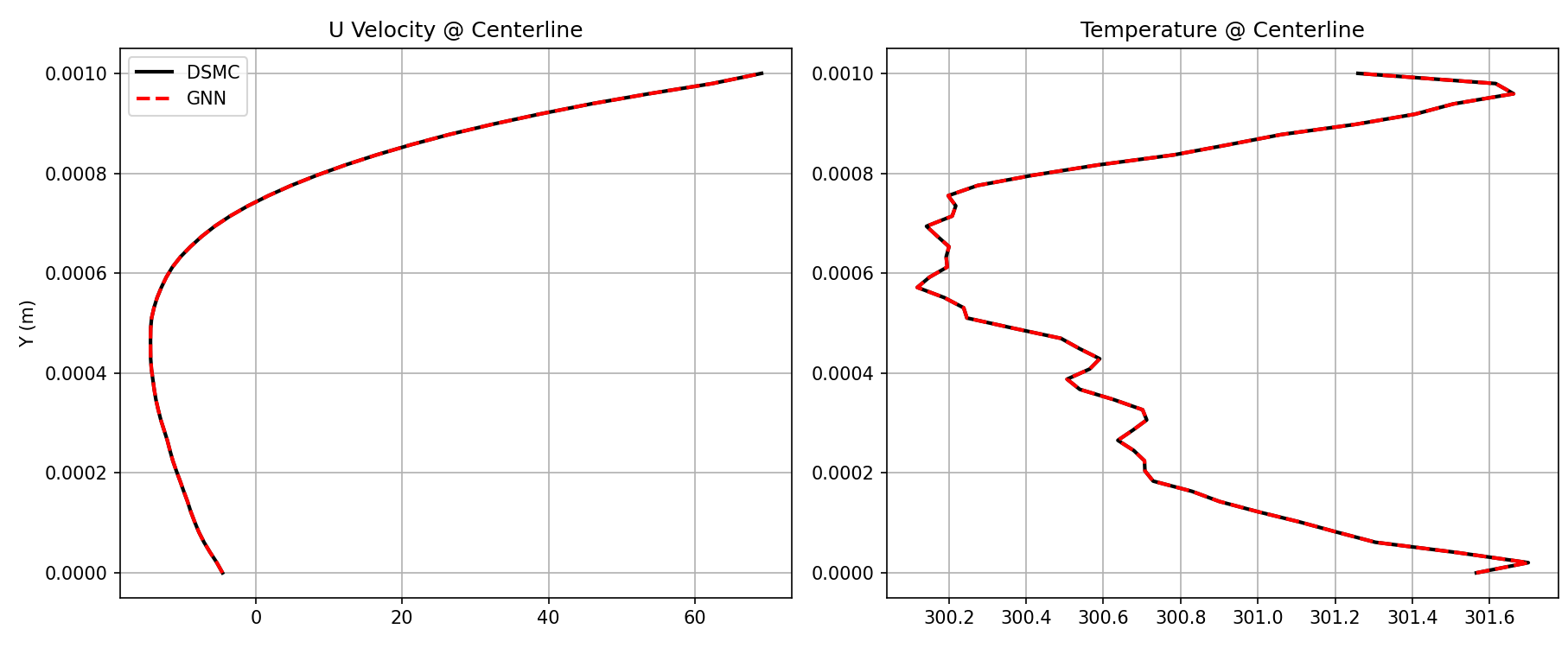}
    \caption{Quantitative validation along the vertical centerline ($x=L/2$).}
    \label{fig:cavity_centerline_primary}
\end{figure*}

\begin{figure*}[htbp]
    \centering
    \includegraphics[width=1.0\textwidth]{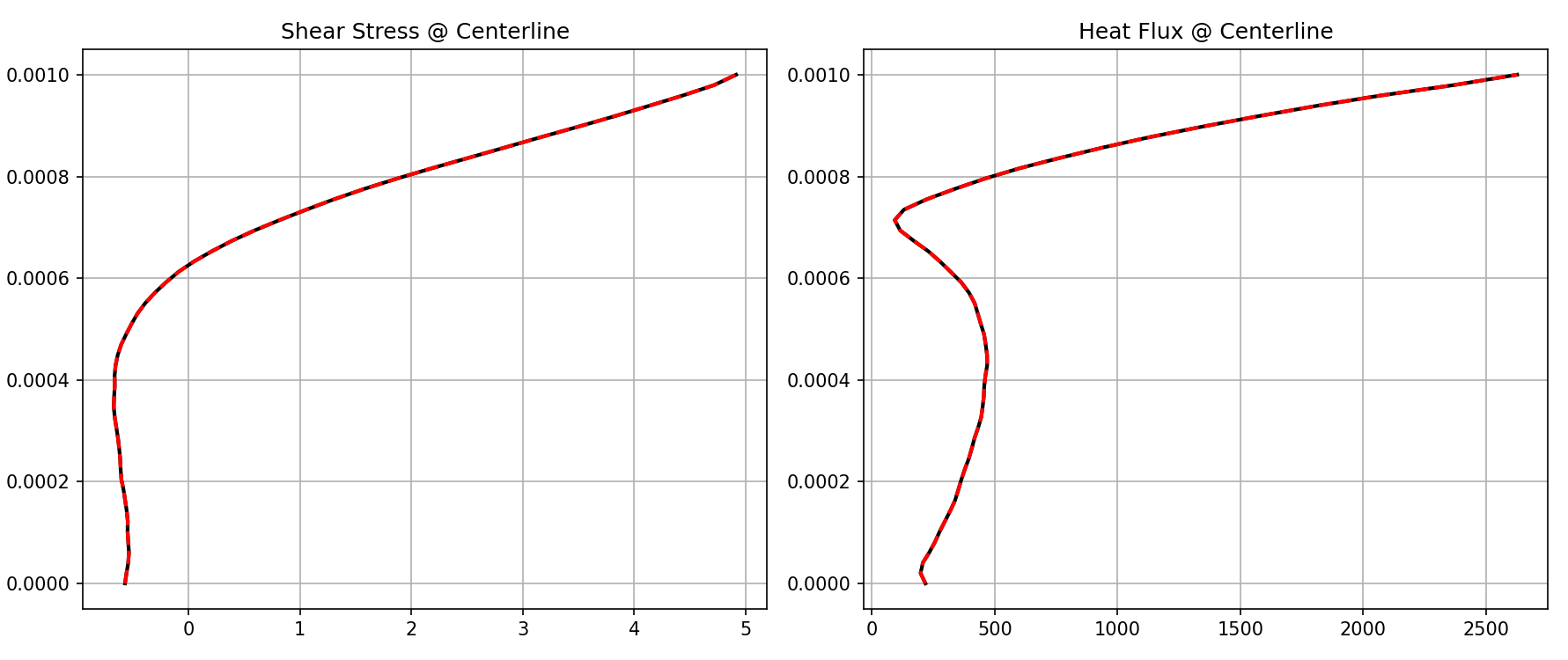}
    \caption{Quantitative validation of higher-order moments along the vertical centerline.}
    \label{fig:cavity_centerline_higher}
\end{figure*}

Remarkably, the agreement extends beyond the primary variables ($U, T$) to the higher-order moments of the distribution function. As shown in Figure \ref{fig:cavity_higher_order}, the shear stress ($\tau_{xy}$) and heat flux ($q$) contours matches the DSMC benchmark with high precision. This indicates that the Moment Matching layer effectively conserves the transport properties of the gas even in multidimensional flows.

To provide a quantitative validation, Figures \ref{fig:cavity_centerline_primary}-\ref{fig:cavity_centerline_higher} present the profiles of velocity, temperature, shear stress, and heat flux along the vertical centerline of the cavity ($x=L/2$).

The centerline profiles confirm that the DNNcaptures the velocity slip at the moving wall ($y=1.0$ mm) and the temperature jump boundary conditions, which are critical in the slip flow regime. The overlap of the shear stress and heat flux profiles further demonstrates that the neural operator correctly models the constitutive relations of the rarefied gas, validating its potential as a geometry-independent surrogate for the collision integral.

The probability density function (PDF) of the scattering angle cosine ($\cos \chi$) predicted by the DNN (red dashed line) is compared against the ground-truth DSMC statistics (black solid line) for the 2D Cavity flow. The suitable overlap confirms that the DNN, despite being trained on 1D Couette flow, correctly reproduces the isotropic scattering law (characteristic of the VHS model) in a completely different flow geometry, validating its physical generalization capability. It is observed that both the DSMC and DNNscattering distributions exhibit minor statistical fluctuations around the theoretical isotropic value. These fluctuations are inherent to the stochastic nature of the Monte Carlo sampling process with a finite sample size ($N \approx 10^4$). The key finding is that the DNN surrogate accurately reproduces the statistical properties (mean and variance) and the overall shape of the PDF, demonstrating that it acts as a physically consistent stochastic generator rather than a deterministic function.

\begin{figure}[htbp]
    \centering
    \includegraphics[width=0.8\textwidth]{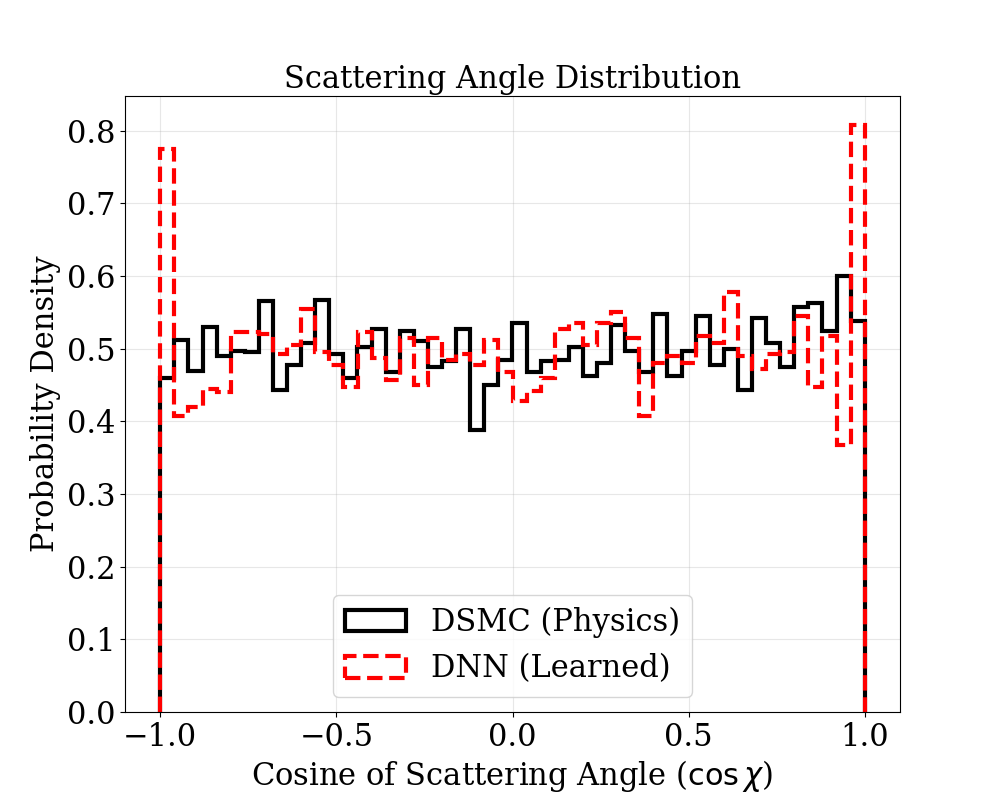}
    \caption{Microscopic validation of the learned collision operator.}
    \label{fig:scattering_validation}
\end{figure}

\subsection{Ab Initio Collision Dynamics and Machine Learning Surrogate}

\subsubsection{The Jager \textit{ab initio} Potential for Argon}
To ensure high-fidelity simulation of inter-atomic interactions beyond the limitations of the Variable Hard Sphere (VHS) or Variable Soft Sphere (VSS) models, we employ the state-of-the-art \textit{ab initio} potential for Argon-Argon interactions proposed by Jager et al.~\cite{jager2009interatomic}. This potential represents a modified Tang-Toennies model, designed to accurately capture both the steep short-range repulsion and the long-range dispersion forces, which are critical for resolving transport properties across a wide temperature regime.

The potential energy surface, $V(r)$, as a function of the inter-atomic distance $r$, is defined by a hybrid function to avoid singularities at extremely short distances. For the standard interaction range ($r \ge 0.4 R_{\epsilon}$), the potential is given by a combination of a Born-Mayer repulsive term and damped dispersion attraction:

\begin{equation}
    V(r) = A \exp\left( a_1 r + a_2 r^2 + \frac{a_{-1}}{r} + \frac{a_{-2}}{r^2} \right) - \sum_{n=3}^{8} f_{2n}(br) \frac{C_{2n}}{r^{2n}},
    \label{eq:jager_potential}
\end{equation}

where $A$ is the repulsive amplitude, and the coefficients $a_i$ govern the shape of the repulsive wall. The dispersion coefficients $C_{2n}$ (specifically $C_6, C_8, \dots, C_{16}$) account for multipole interactions. To prevent unphysical divergence of the attractive terms at small $r$, the Tang-Toennies damping function, $f_n(x)$, is applied:

\begin{equation}
    f_n(x) = 1 - \exp(-x) \sum_{k=0}^{n} \frac{x^k}{k!},
    \label{eq:damping}
\end{equation}

where $x = b \cdot r$. For extremely short distances ($r < 0.4 R_{\epsilon}$), where the theoretical expansion may fail, the potential is regularized using a screened Coulomb form:

\begin{equation}
    V_{short}(r) = \frac{\tilde{A}}{r} \exp(-\tilde{a} r).
\end{equation}

The specific coefficients used in this study ($A, b, C_n$, etc.) are adopted directly from the spectroscopic parameters provided by Jager et al., ensuring that the collision dynamics reflect quantum-mechanical accuracy.

\subsubsection{Neural Network Surrogate Model for Scattering Angles}
Direct integration of the scattering angle $\chi$ for every collision event in a DSMC solver is computationally prohibitive due to the complexity of Eq. (\ref{eq:jager_potential}). To address this, we developed a Deep Neural Network (DNN) surrogate model trained to map the collision parameters directly to the scattering angle. The schematic of the employed neural network is shown in Fig.~\ref{fig:nn_architecture}.

\subsubsection*{Data Generation and Architecture}
The training dataset was generated by numerically solving the classical scattering integral using a parallelized CPU architecture (50 cores). The exact scattering angle $\chi$ is calculated as:
\begin{equation}
    \chi(b, E_r) = \pi - 2b \int_{r_{min}}^{\infty} \frac{dr}{r^2 \sqrt{1 - (b/r)^2 - V(r)/E_r}},
\end{equation}
where $b$ is the impact parameter, $E_r$ is the relative collision energy, and $r_{min}$ is the distance of closest approach. We generated over 200,000 samples spanning relative energies corresponding to temperatures from $5\,\text{K}$ to $60,000\,\text{K}$ and impact parameters up to a cutoff of $b_{max} = 1.2\,\text{nm}$.

 \tikzset{input/.style={circle, draw, fill=blue!20, minimum size=20pt, inner sep=0pt, text centered}}
 \tikzset{arrow/.style={->,thick}}
\begin{figure}[htbp]
    \centering
    \begin{tikzpicture}[
        shorten >=1pt,->,
        draw=black!50,
        node distance=2.2cm,
        neuron/.style={circle, draw=black!100, minimum size=25pt, inner sep=0pt, thick},
        input/.style={neuron, fill=green!20},
        hidden/.style={neuron, fill=blue!20},
        output/.style={neuron, fill=red!20},
        annot/.style={text width=6em, align=center}
    ]

        \foreach \name / \y / \text in {1/1.5/$\ln(E_r)^*$, 2/3.5/$b/b_{max}$}
        \foreach \name/\y/\lbl in {
          1/1.5/{$\ln(E_r)^{*}$},
          2/3.5/{$b/b_{max}$}
        }{
          \node[input] (I-\name) at (0,-\y) {\small \lbl};
        }

        \node[annot, above of=I-1, node distance=1.5cm] (hl-in) {\textbf{Input Layer}};
        \node[annot, below of=I-2, node distance=1.5cm] {2 Neurons};

        \node[hidden] (H1-1) at (2.5,0) {};
        \node[hidden] (H1-2) at (2.5,-1) {};
        \node[hidden] (H1-3) at (2.5,-2) {};
        \node (H1-dots) at (2.5,-3) {$\vdots$};
        \node[hidden] (H1-4) at (2.5,-5) {};

        \node[annot, above of=H1-1, node distance=1.5cm] (hl-1) {\textbf{Hidden 1}};
        \node[annot, below of=H1-4, node distance=1.5cm] {512 Neurons\\(SiLU)};

        \node[hidden] (H2-1) at (5,0) {};
        \node[hidden] (H2-2) at (5,-1) {};
        \node[hidden] (H2-3) at (5,-2) {};
        \node (H2-dots) at (5,-3) {$\vdots$};
        \node[hidden] (H2-4) at (5,-5) {};

        \node[annot, above of=H2-1, node distance=1.5cm] (hl-2) {\textbf{Hidden 2}};
        \node[annot, below of=H2-4, node distance=1.5cm] {512 Neurons\\(SiLU)};

        \node[hidden] (H3-1) at (7.5,0) {};
        \node[hidden] (H3-2) at (7.5,-1) {};
        \node[hidden] (H3-3) at (7.5,-2) {};
        \node (H3-dots) at (7.5,-3) {$\vdots$};
        \node[hidden] (H3-4) at (7.5,-5) {};

        \node[annot, above of=H3-1, node distance=1.5cm] (hl-3) {\textbf{Hidden 3}};
        \node[annot, below of=H3-4, node distance=1.5cm] {512 Neurons\\(SiLU)};

        \node[hidden] (H4-1) at (10,-0.5) {};
        \node[hidden] (H4-2) at (10,-1.5) {};
        \node (H4-dots) at (10,-2.5) {$\vdots$};
        \node[hidden] (H4-3) at (10,-4.5) {};

        \node[annot, above of=H4-1, node distance=1.5cm] (hl-4) {\textbf{Hidden 4}};
        \node[annot, below of=H4-3, node distance=1.5cm] {256 Neurons\\(SiLU)};

        \node[hidden] (H5-1) at (12.5,-1) {};
        \node (H5-dots) at (12.5,-2.5) {$\vdots$};
        \node[hidden] (H5-2) at (12.5,-4) {};

        \node[annot, above of=H5-1, node distance=2cm] (hl-5) {\textbf{Hidden 5}};
        \node[annot, below of=H5-2, node distance=1.5cm] {128 Neurons\\(SiLU)};

        \node[output] (O-1) at (15,-2.5) {$\chi$};

        \node[annot, above of=O-1, node distance=3.5cm] (hl-out) {\textbf{Output}};
        \node[annot, below of=O-1, node distance=2.5cm] {1 Neuron\\(Softplus)};

        \foreach \source in {1,2}
            \foreach \dest in {1,2,3,4}
                \draw[arrow] (I-\source) -- (H1-\dest);

        \foreach \source in {1,2,3,4}
            \foreach \dest in {1,2,3,4}
                \draw[arrow] (H1-\source) -- (H2-\dest);

        \foreach \source in {1,2,3,4}
            \foreach \dest in {1,2,3,4}
                \draw[arrow] (H2-\source) -- (H3-\dest);

        \foreach \source in {1,2,3,4}
            \foreach \dest in {1,2,3}
                \draw[arrow] (H3-\source) -- (H4-\dest);

        \foreach \source in {1,2,3}
            \foreach \dest in {1,2}
                \draw[arrow] (H4-\source) -- (H5-\dest);

        \foreach \source in {1,2}
            \draw[arrow] (H5-\source) -- (O-1);

    \end{tikzpicture}
    \caption{Schematic representation of the Neural Network architecture used for scattering angle prediction. The network consists of 5 hidden layers with decreasing width in the deeper layers. SiLU activation is used for hidden layers and Softplus for the output to ensure positivity.}
    \label{fig:nn_architecture}
\end{figure}
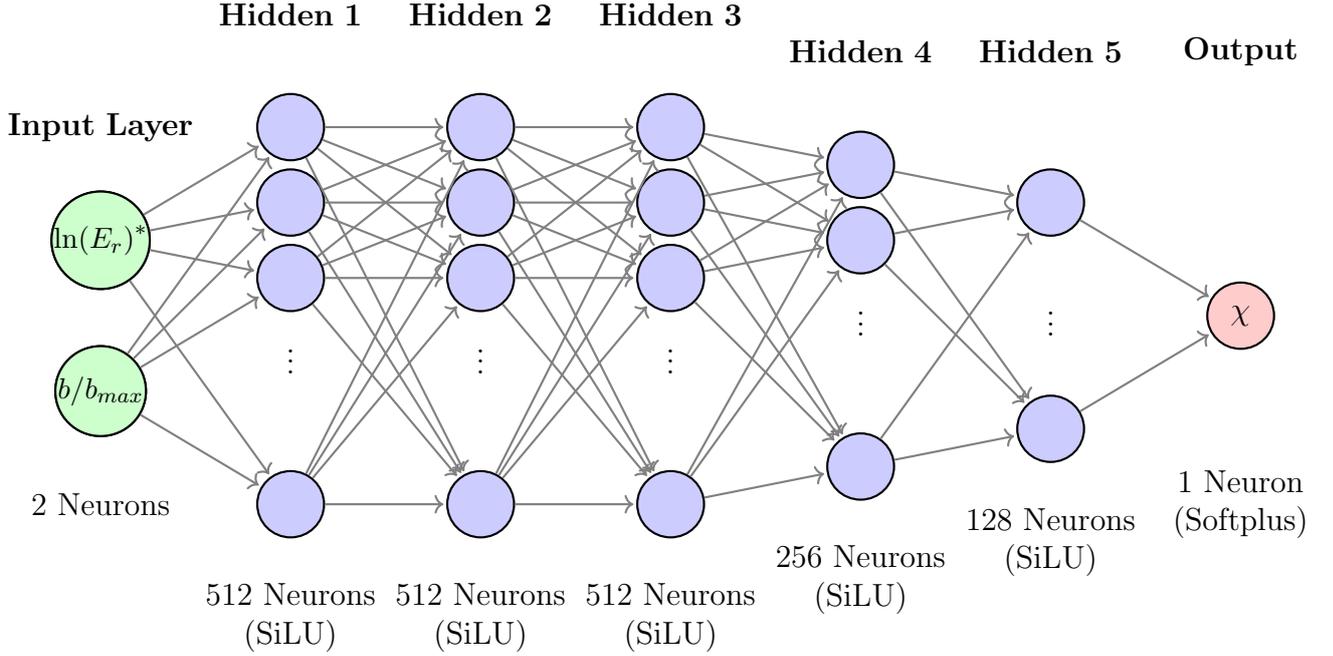

The mathematical operation for each layer $l$ is defined as $\boldsymbol{h}^{(l)} = \sigma(\boldsymbol{W}^{(l)} \boldsymbol{h}^{(l-1)} + \boldsymbol{b}^{(l)})$, where the SiLU (Sigmoid Linear Unit) activation function, defined as $\sigma(z) = z / (1 + e^{-z})$, was chosen for its smoothness and non-monotonicity, which aids in capturing the "rainbow scattering" peaks observed in the physical model. The network was trained using the Adam optimizer to minimize the Mean Squared Error (MSE) of the logarithmically transformed output, $\ln(1+\chi)$, ensuring high accuracy across dynamic ranges.

\subsubsection*{Microscopic Validation}
\begin{figure}[htbp]
    \centering
    \includegraphics[width=1.0\textwidth]{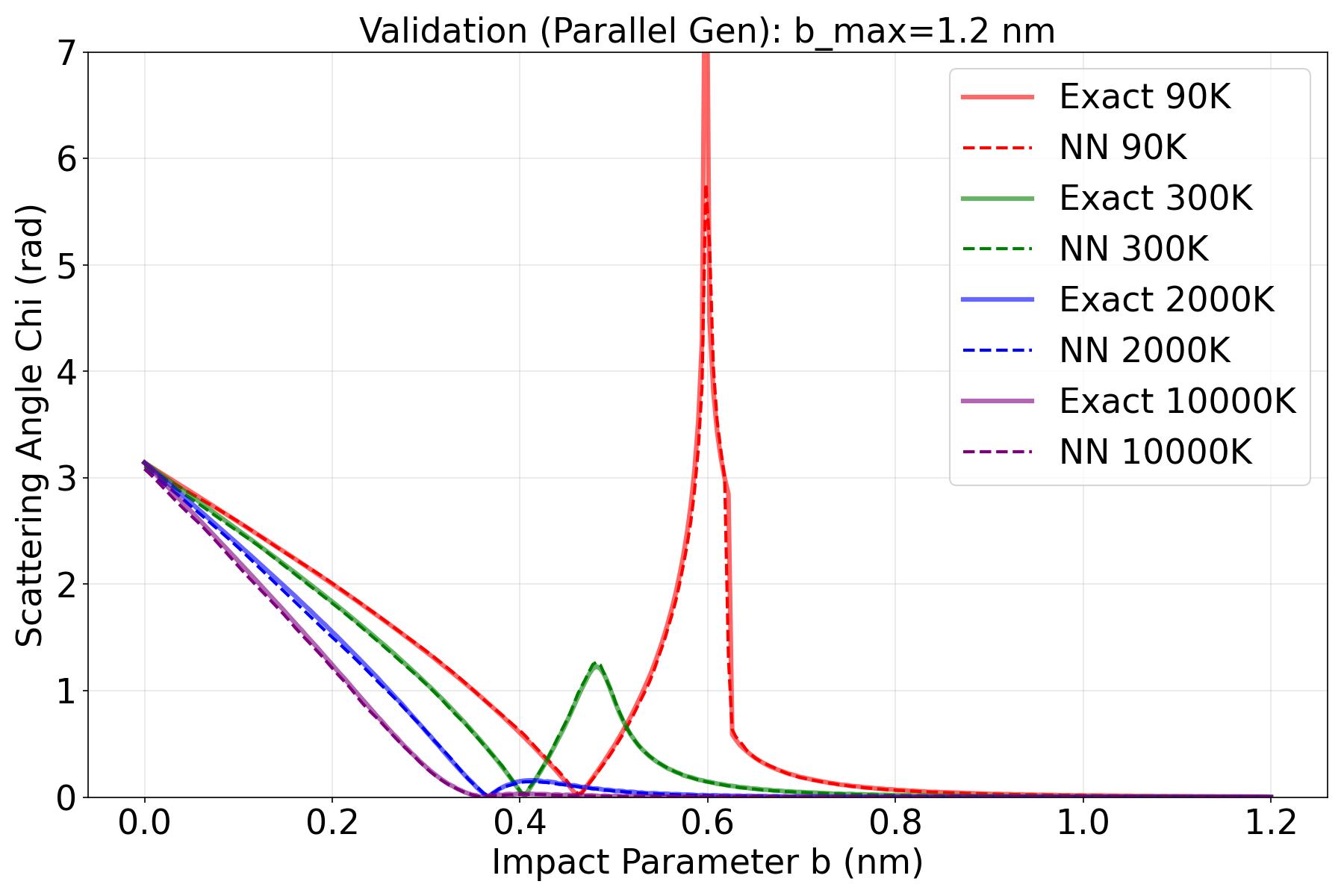}
    \caption{Microscopic verification of the Neural Network surrogate model against the exact numerical solution of the Jager potential. The scattering angle $\chi$ is plotted as a function of the impact parameter $b$ for various relative energies corresponding to temperatures of 90 K, 300 K, 2000 K, and 10,000 K. The dashed lines (Neural Network predictions) suitably overlay the solid lines (Exact Physics), confirming the fidelity of the surrogate model up to $b_{max} = 1.2$ nm.}
    \label{fig:nn_validation}
\end{figure}

Figure \ref{fig:nn_validation} demonstrates the performance of the trained neural network against the exact numerical integration of the Jager potential. The results show agreement across varying energy regimes. Notably, the network accurately captures the orbiting phenomenon at low temperatures (e.g., 90 K), characterized by sharp peaks in the deflection angle, as well as the smooth monotonic decay at high temperatures (e.g., 10,000 K). The profiles shown in Fig.~\ref{fig:nn_validation} are similar to profiles plotted in Fig. 1 of Ref.~\cite{sharipov2012abinitio}.

\begin{figure}[htbp]
    \centering
    \includegraphics[width=0.98\textwidth]{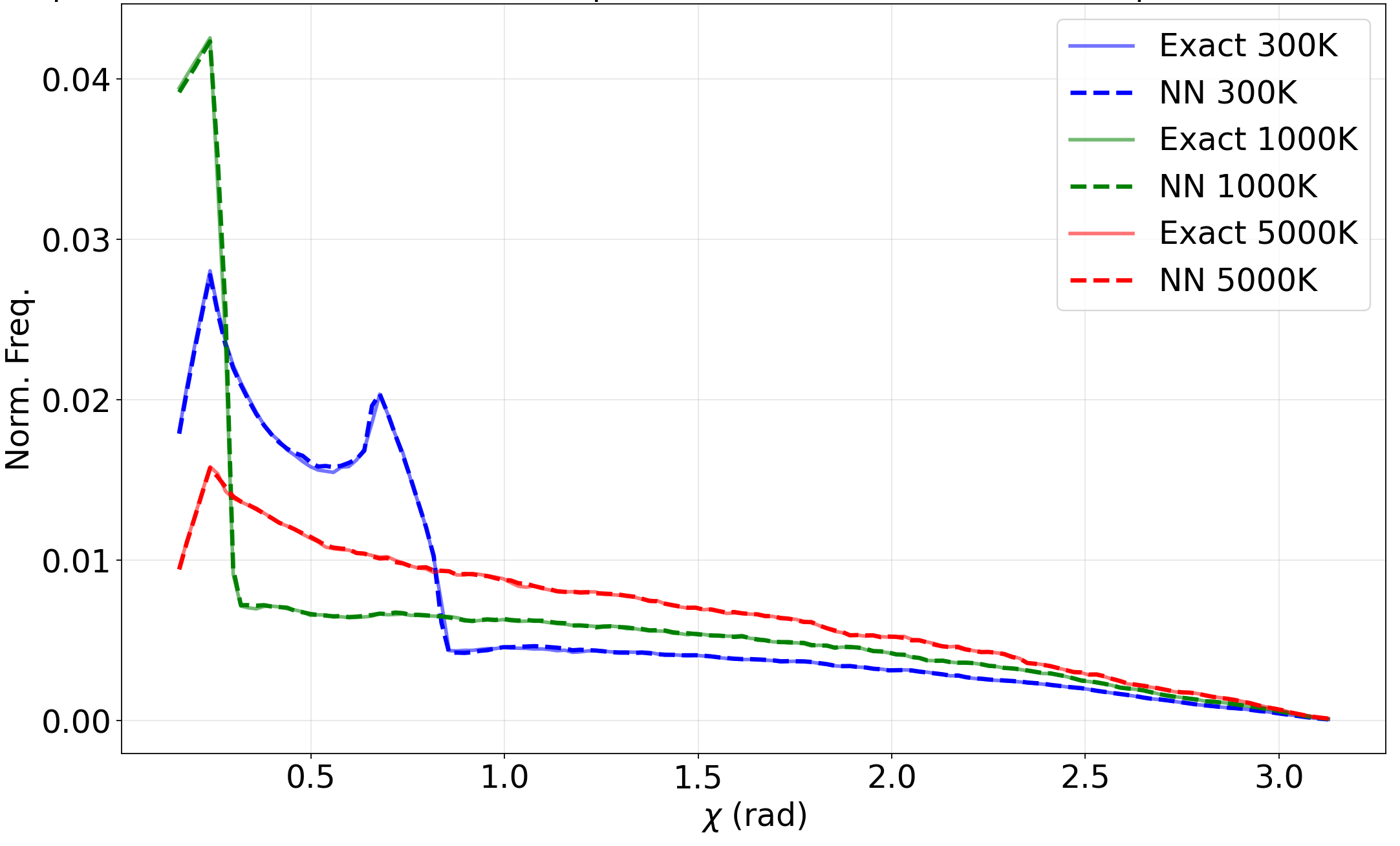}
    \caption{Distribution-level validation of the neural surrogate for the deflection angle $\chi$ at 300~K, 1000~K, and 5000~K. Solid lines show the exact physics and dashed lines show neural predictions. Exact and neural samples are compared using a common masking range and identical histogram bins to ensure a consistent PDF comparison.}
    \label{fig:chi_pdf_validation}
\end{figure}

While Fig.~\ref{fig:nn_validation} verifies the pointwise accuracy of the surrogate through $\chi(b)$ curves, DSMC requires the \emph{distribution} of scattering outcomes to be reproduced correctly under physically relevant sampling of the impact parameter. Therefore, we additionally compare the probability density of deflection angles $\chi$ generated by the neural surrogate against the exact physics at representative temperatures (300~K, 1000~K, and 5000~K). To avoid bias from different clipping or truncation between datasets, the exact and neural samples are evaluated using a common masking range and identical histogram bins. As shown in Fig.~\ref{fig:chi_pdf_validation}, the neural surrogate reproduces the full $\chi$-distribution across temperatures, including the low-$\chi$ plateau and the high-$\chi$ tail, confirming that the model is suitable for Monte Carlo collision sampling in hypersonic regimes.

\subsubsection{Macroscopic Validation: Viscosity and Transport Coefficients}
To validate the macroscopic implications of the implemented interaction model, we calculated the shear viscosity of Argon and compared it with experimental benchmarks. According to the Chapman-Enskog theory of dilute gases, the first-order approximation for viscosity, $\mu$, is related to the collision integral $\Omega^{(2,2)}$ by:

\begin{equation}
    \mu(T) = \frac{5}{16} \frac{\sqrt{\pi m k_B T}}{\bar{\Omega}^{(2,2)}(T)},
    \label{eq:chapman_enskog}
\end{equation}

where $m$ is the atomic mass and $k_B$ is the Boltzmann constant. The collision integral is obtained by averaging the transport cross-section $Q^{(2)}(E)$ over a Maxwellian energy distribution:
\begin{equation}
    \bar{\Omega}^{(2,2)}(T) = \frac{1}{2 (k_B T)^3} \int_0^{\infty} E_r^2 Q^{(2)}(E_r) \exp\left(-\frac{E_r}{k_B T}\right) dE_r,
\end{equation}
with the transport cross-section defined as:
\begin{equation}
    Q^{(2)}(E_r) = 2\pi \int_0^{\infty} (1 - \cos^2 \chi) b \, db.
\end{equation}

\begin{figure}[htbp]
    \centering
    \includegraphics[width=1.0\textwidth]{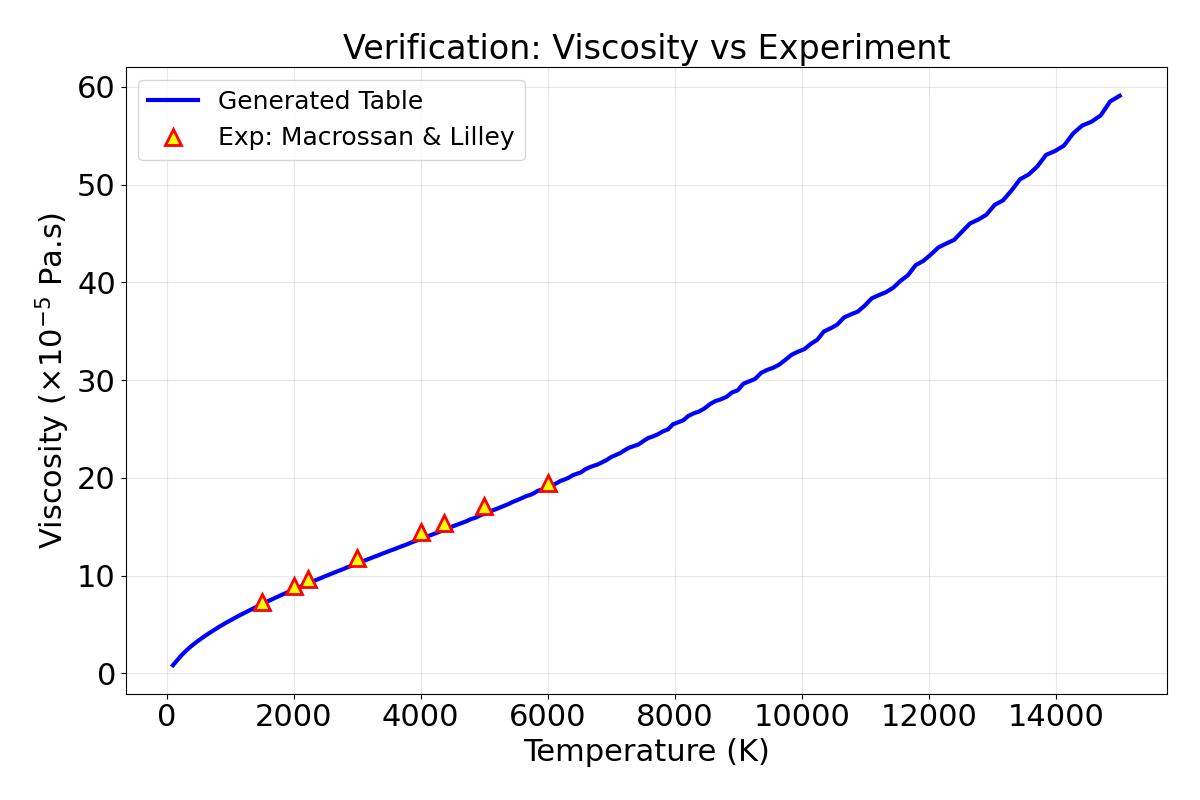}
    \caption{Macroscopic validation of the collision model. The solid blue line represents the viscosity calculated using the Chapman-Enskog theory based on the generated Jager potential table. The red triangles represent the experimental data points and correlations from Macrossan \& Lilley (2003). The excellent agreement validates the accuracy of the underlying interaction potential and the table generation process.}
    \label{fig:viscosity_check}
\end{figure}

We computed these integrals numerically using the Jager potential. Figure \ref{fig:viscosity_check} presents the comparison between our calculated viscosity (utilizing the generated look-up table) and the high-temperature experimental data provided by Macrossan and Lilley~\cite{macrossan2003viscosity}.

The results indicate a precise alignment between our \textit{ab initio} derived viscosity and the established experimental data across the temperature range of 1,500 K to 6,000 K. This confirms that the microscopic collision model, encapsulated within the neural network, correctly reproduces the macroscopic transport properties of the gas.

\subsubsection{Computational Implementation: The Exact vs. Neural Paradigm}

To rigorously quantify the efficiency and fidelity of the proposed method in hypersonic regimes, we developed two distinct collision kernels within the DSMC solver. The first establishes a ground truth using direct numerical integration of the quantum-mechanical potential, while the second utilizes the proposed neural surrogate via a high-performance look-up strategy.

The "Exact" solver operates by resolving the classical scattering integral for every unique collision pair (or grid point during table initialization) derived from the Jager potential $V(r)$. The scattering angle $\chi$ is governed by the conservation of angular momentum and energy:
\begin{equation}
    \chi(b, E_r) = \pi - 2b \int_{r_{min}}^{\infty} \frac{dr}{r^2 \sqrt{1 - \frac{b^2}{r^2} - \frac{V(r)}{E_r}}}
\end{equation}
where $E_r = \frac{1}{2}\mu v_r^2$ is the relative collision energy. 

The computational bottleneck of this approach is twofold:
The lower limit of integration, $r_{min}$ (distance of closest approach), is defined as the largest root of the denominator:
    \begin{equation}
        1 - \frac{b^2}{r_{min}^2} - \frac{V(r_{min})}{E_r} = 0
    \end{equation}
    Since $V(r)$ in the Jager model (Eq. 12) is a complex combination of exponential repulsion and damped dispersion terms, Eq. (19) is transcendental. We employ the Brent-Dekker method (`scipy.optimize.brentq`) to solve for $r_{min}$ with a tolerance of $10^{-14}$. This iterative search must be performed for every single query, consuming significant CPU cycles.
    
The integrand in Eq. (19) possesses a singularity at $r = r_{min}$. To handle this, we utilize the substitution $u = r_{min}/r$, transforming the integral domain to $[0, 1]$ and removing the singularity analytically. The transformed integral is then solved using the adaptive Gauss-Kronrod quadrature (`scipy.integrate.quad`).

A natural question is why a neural network is needed if the DSMC solver ultimately samples from a look-up table. The key point is that, for ab initio interactions, the table itself is the expensive object: constructing a high-resolution table (e.g., 512×512) via direct evaluation of the scattering integral requires solving 260,000 coupled root-finding and singular quadrature problems, because the distance of closest approach must be obtained from a transcendental equation and the scattering integral exhibits a near-singularity at $r = r_{\min}$. This makes the “Exact” pathway a severe pre-computation bottleneck when generated on CPUs and/or when repeated for multiple temperature bands, mixtures, or updated potentials.

From a computational hardware perspective, the neural-generated ab initio table offers significant advantages. A high-resolution $512 \times 512$ scattering table occupies approximately 1.0 to 2.0 MB of VRAM, a memory footprint small enough to reside almost entirely within the L2 or L3 cache of modern GPU/CPU architectures. This ensures that the collision kernel achieves true O(1) complexity with near-zero memory latency, bypassing the expensive transcendental root-finding and singular quadrature evaluations required by the "Exact" approach. This hardware-aware optimization transforms ab initio DSMC from a pre-computationally heavy task into a high-throughput, constant-time memory fetch operation.

The neural surrogate eliminates this bottleneck by enabling on-the-fly table generation: for any user-specified collision-energy range (or for new mixtures/potentials that require re-tabulation), the network can synthesize the full $\chi$-table directly on the GPU in a negligible wall time, after which DSMC samples collisions through constant-time memory fetches. This design decouples high-fidelity scattering physics from expensive numerical integration, turning the cost of \textit{ab initio} scattering from ``hours of pre-processing'' into ``seconds of GPU inference,'' and thus makes re-tabulation practical in parametric studies and multi-condition hypersonic workflows.

In addition, the learned mapping acts as a regularized physics filter: numerical integration in the low-energy orbiting/rainbow regime can introduce high-frequency artifacts or non-physical waviness in $\chi$, whereas the surrogate---trained with robust objectives and annealed optimization---suppresses such numerical noise and yields a smoother, more stable table for repeated DSMC sampling.

\subsubsection*{Neural Surrogate Training: The "Physics Harvesting" Strategy}
To circumvent the cost of numerical integration while maintaining \textit{ab initio} accuracy, the Neural Network was trained using a specialized strategy tailored for the high-gradient shock regime.

\textbf{Dataset Generation:} A massive dataset of $\mathcal{D} = 2.5 \times 10^6$ collision pairs was generated using parallel CPU processing. The sampling strategy was designed to cover the extreme conditions of Mach 10 flow:
Relative energies were sampled logarithmically from $T \approx 5$ K to $T = 60,000$ K. This ensures the network captures both the low-energy "orbiting" resonances (Rainbow scattering) and the high-energy repulsive core collisions dominant in the bow shock.
The impact parameter $b$ was sampled with a bias towards head-on collisions ($b \to 0$) to accurately resolve the high-angle scattering events that govern viscosity in the shock layer.

A critical challenge in learning \textit{ab initio} potentials is ensuring the monotonicity of $\chi$ at high energies. Standard Mean Squared Error (MSE) loss often induces non-physical "waviness" in the function approximation at $T > 5000$ K. To resolve this, we employed the Smooth L1 Loss function (Huber Loss), defined as:
\begin{equation}
    \mathcal{L}(y, \hat{y}) = \frac{1}{N} \sum_{i=1}^{N} \begin{cases} 
      0.5 (y_i - \hat{y}_i)^2, & \text{if } |y_i - \hat{y}_i| < 1 \\
      |y_i - \hat{y}_i| - 0.5, & \text{otherwise}
   \end{cases}
\end{equation}
where $y = \ln(1+\chi)$ is the target. This loss function is robust to outliers in the orbiting regime while enforcing smoothness in the repulsive regime. Furthermore, we utilized a Cosine Annealing Learning Rate Scheduler (via \path{torch.optim.lr\_scheduler.CosineAnnealingLR}), which decays the learning rate to near-zero ($10^{-6}$) in the final epochs. This technique allows the optimizer to settle into a wide, flat local minimum, effectively eliminating high-frequency numerical noise from the predicted scattering curves.

\subsubsection*{Runtime Acceleration via Neural Table Look-up}
In the final hypersonic simulation, direct neural inference for every collision pair ($N_{coll} > 10^9$ over the simulation lifetime) remains computationally demanding on GPU tensor cores. Therefore, we adopt a hybrid "Neural Look-up" strategy that decouples inference from the time-stepping loop:

Upon initialization, the trained Compact-MLP (6 layers) is queried to populate a dense scattering angle table ($512 \times 512$ grid). This inference step takes less than 1.0 second on a single GPU, compared to several hours required for the Exact integration method on a CPU cluster.
    
The table is parameterized by normalized impact parameter $\tilde{b} = b/b_{max}$ and logarithmic energy $\tilde{E} = \ln(E_r/k_B)$. The logarithmic energy scale is crucial for handling the orders-of-magnitude variation in collision energies across the shock wave.
    
During the DSMC collision kernel execution, the solver does not compute the potential or run the neural network. Instead, it calculates the collision parameters $(\tilde{b}, \tilde{E})$ and retrieves the scattering angle $\chi$ via bilinear interpolation from the pre-loaded GPU memory. This reduces the complex quantum-mechanical interaction to a constant-time $O(1)$ memory access operation.

This methodology guarantees that the simulation benefits from the physical fidelity of the Jager potential (verified by the accurate overlay of scattering distributions in Fig. 7) while maintaining the computational efficiency required for large-scale engineering simulations.

\subsection{Results: Hypersonic Flow over a Cylinder}

While the Couette and Cavity simulations served primarily to validate the learning accuracy of the surrogate model for a simple VHS model without focusing on performance gains, the true computational advantage of the proposed method is realized in complex interaction models. This section introduces the hypersonic cylinder flow using \textit{ab initio} collision dynamics, where the surrogate model yields a substantial reduction in computational cost by bypassing the expensive scattering calculations.
Hypersonic rarefied flow over a cylinder serves as a fundamental benchmark for validating kinetic solvers and has been extensively investigated in the literature \cite{lofthouse2008effects, emerson2012assessment, volkov2012direct}.
While earlier studies predominantly utilized phenomenological collision models, Volkov and Sharipov \cite{volkov2012direct} advanced the field by applying \textit{ab initio} potentials to this canonical problem, demonstrating the importance of accurate intermolecular forces in high-speed regimes.
The implementation of arbitrary intermolecular potentials within the DSMC framework was pioneered by Sharipov and Strapasson \cite{sharipov2012direct, sharipov2012abinitio}. 
They introduced a computationally efficient scheme based on pre-calculated deflection angle tables, demonstrating that \textit{ab initio} potentials could be utilized with a computational cost comparable to the standard Variable Hard Sphere (VHS) model. 
This methodology effectively removed the reliance on empirical fitting parameters for noble gases, enabling high-fidelity simulations of transport phenomena over a wide range of temperatures \cite{sharipov2012abinitio}.
In the present study, the \textit{ab initio} potential data for Argon, specifically the interaction energies and scattering angles, were adopted from the high-fidelity calculations provided by Jager et al. \cite{jager2009interatomic}.

\begin{figure}[htbp]
    \centering
    \includegraphics[width=0.8\textwidth]{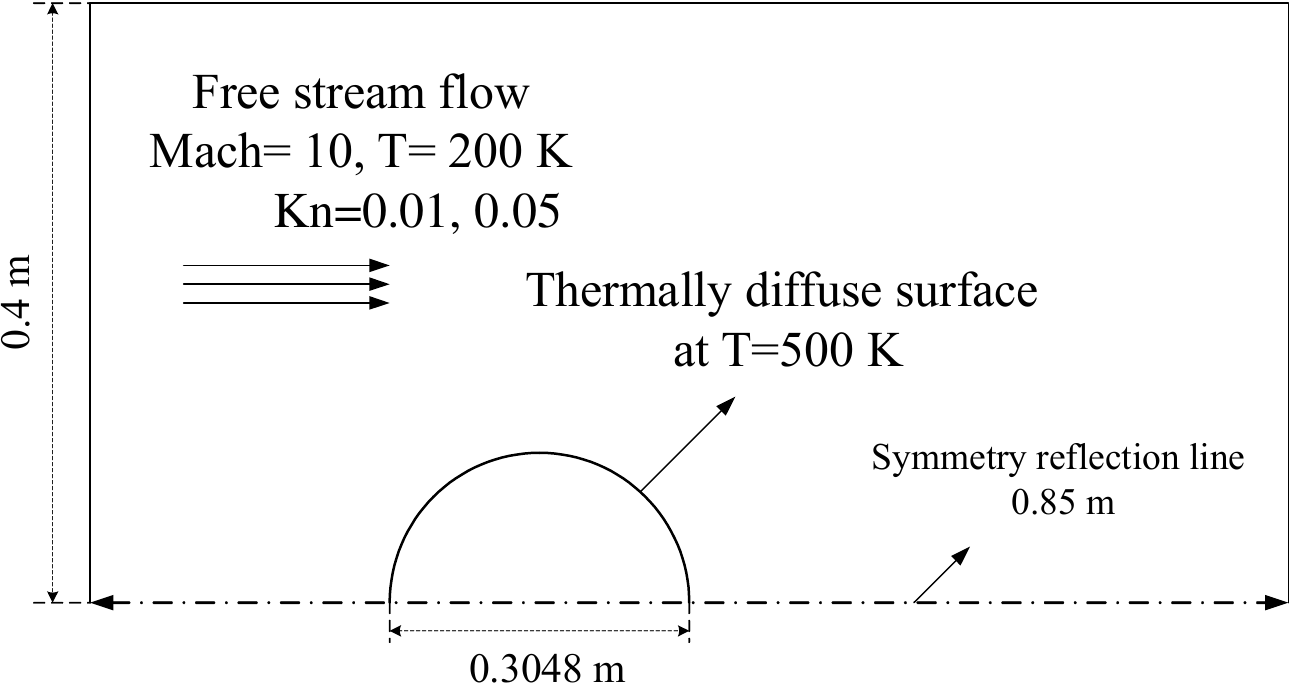}
    \caption{Schematic of the computational domain for hypersonic flow over a cylinder at Mach 10. Kn=0.05 was considered for validation, Kn=0.01 was considered as the main test case.}
    \label{fig:cyl_schematic}
\end{figure}

The predictive capability of the proposed physics-constrained neural operator was validated by simulating a Mach 10 Argon flow over a cylinder, see Fig.~\ref{fig:cyl_schematic}. The freestream conditions were set to $T_{\infty} = 200$ K, $U_{\infty} = 2634.1$ m/s, and a number density of $n = 4.247 \times 10^{20} m^{-3}$, corresponding to a Knudsen number of $Kn = 0.01$ based on the cylinder radius $R=0.1524$ m. The wall temperature was fixed at $T_w = 1000$ K.
To rigorous benchmarking, we compare the results of the "DSMC ML (Ab-Initio)" solver (utilizing the neural look-up table) against the "DSMC Exact (Ab-Initio)" solver (utilizing direct numerical integration).

\subsubsection{Validation against Volkov and Sharipov case}

To rigorously validate the accuracy of the proposed solver in the high-speed rarefied regime, we compared our results against the benchmark numerical study by Volkov and Sharipov \cite{volkov2012direct}. The test case involves hypersonic Argon gas flow over a circular cylinder, a canonical problem characterized by strong non-equilibrium effects including a detached bow shock and a rarefied wake.

The validation was performed for a specific case corresponding to a free-stream Mach number of $Ma_{\infty} = 10$ and a Knudsen number of $Kn_{\infty} = 0.05$. The physical parameters for this benchmark are defined as Argon (monatomic gas, $m = 6.63 \times 10^{-26}$ kg). Velocity $U_{\infty}$ corresponding to Mach 10, with a free-stream temperature of $T_{\infty} = 200$ K. A fixed isothermal wall temperature of $T_w = 500$ K, resulting in a temperature ratio of $T_w/T_{\infty} = 2.5$.

\begin{figure}[htbp]
    \centering
    \includegraphics[width=0.8\textwidth]{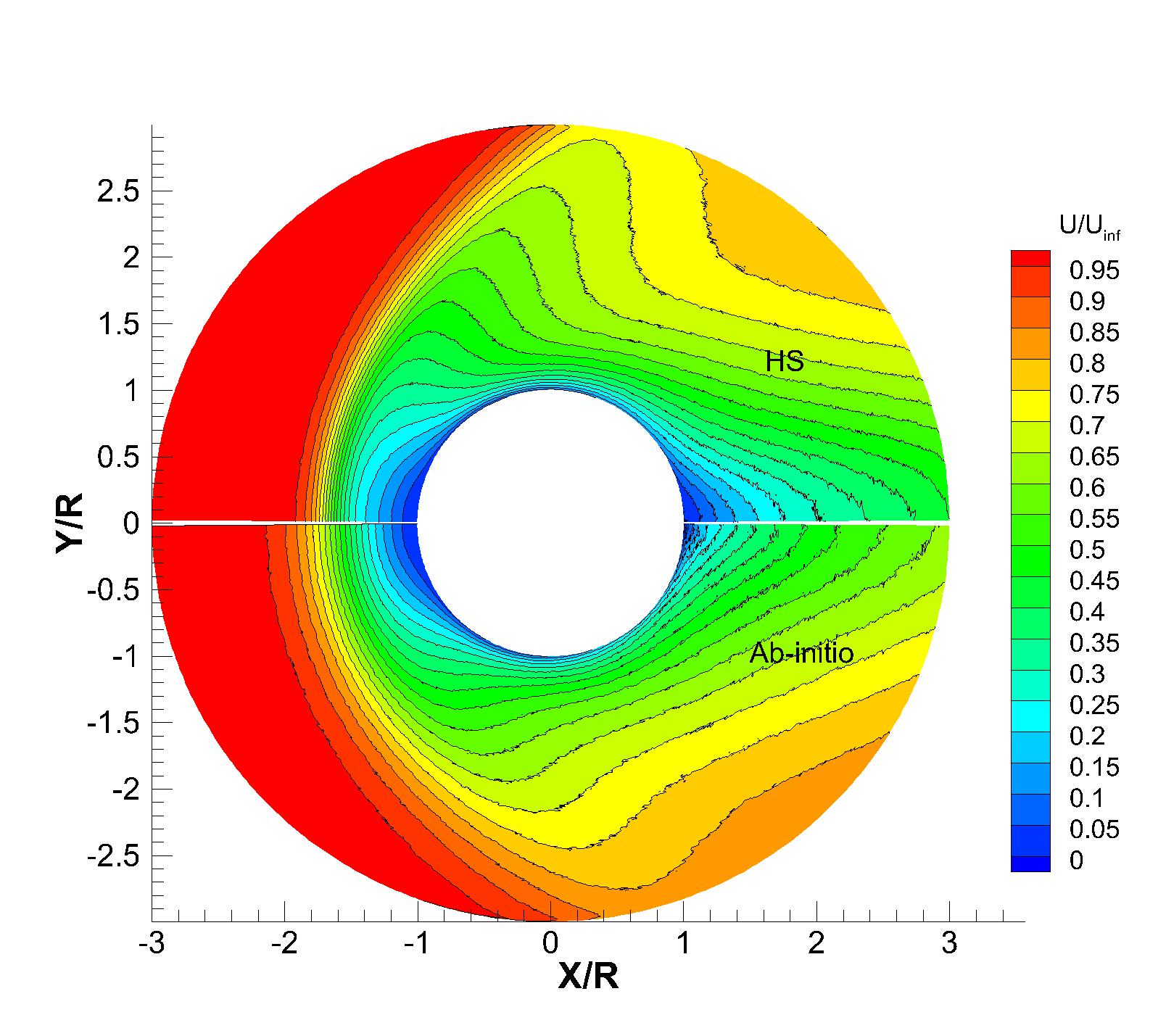}
    \caption{Comparison of normalized velocity magnitude ($U/U_{\infty}$) contours for hypersonic Argon flow at $Ma=10$ and $Kn=0.05$. Top half: Hard Sphere (HS) model. Bottom half: Ab Initio potential. The Ab Initio solution shows a more dispersed shock structure typical of realistic interatomic potentials.}
    \label{fig:contour_U}
\end{figure}

\begin{figure}[h!]
    \centering
    \includegraphics[width=0.8\textwidth]{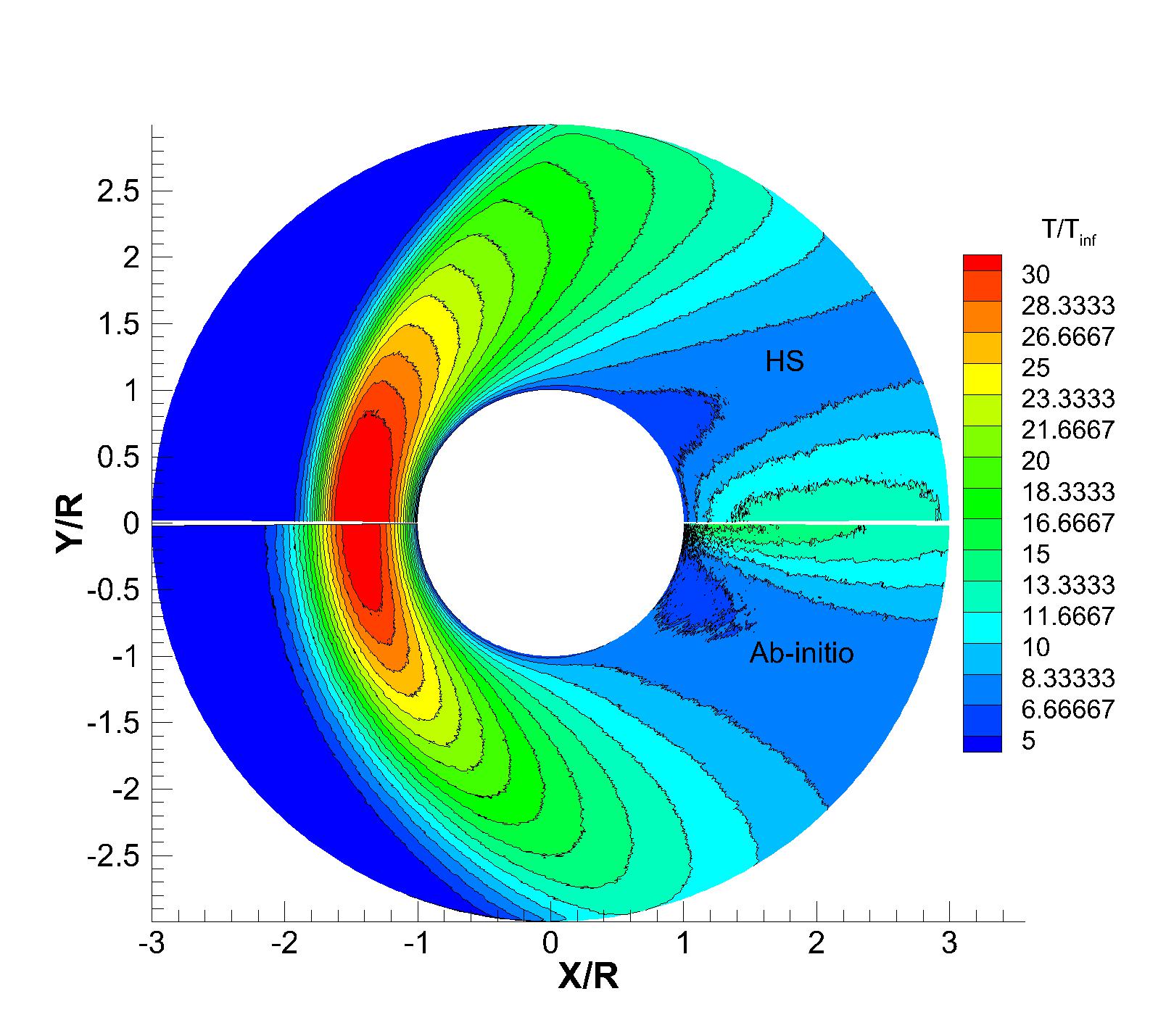}
    \caption{Comparison of normalized translational temperature ($T/T_{\infty}$) contours. Top half: Hard Sphere (HS) model. Bottom half: Ab Initio potential. The thermal gradients in the Ab Initio result are less steep, leading to a larger shock standoff distance consistent with the reference benchmark \cite{volkov2012direct}.}
    \label{fig:contour_T}
\end{figure}

Before comparing the current ab-initio results with Ref.~\cite{volkov2012direct}, we make a comparison between ab-initio and HS predictions of the flow field structure. 
Figures \ref{fig:contour_U} and \ref{fig:contour_T} present the normalized velocity magnitude and temperature contours, respectively. In each figure, the upper half displays the results obtained using the Hard Sphere (HS) model, while the lower half presents the results using the current Ab Initio potential method implemented in our Python script.

A direct comparison reveals distinct differences in the shock wave structure between the two interaction potentials. Consistent with the findings reported in Figure 13 of Volkov and Sharipov \cite{volkov2012direct}, the bow shock predicted by the Ab Initio potential is noticeably more diffuse and exhibits a larger standoff distance from the cylinder body compared to the Hard Sphere result. The Hard Sphere model, which approximates the scattering cross-section more rigidly, tends to predict a steeper shock gradient. Conversely, the Ab Initio potential accurately captures the "soft" nature of the interatomic repulsion at high energies, resulting in a thicker shock layer and a smoother transition from the free stream to the post-shock conditions. The position and topology of the shock wave obtained in our Ab Initio simulation align precisely with the benchmark results provided by Volkov and Sharipov, confirming that the solver correctly resolves the non-equilibrium scattering dynamics within the shock layer.

\begin{figure}[h!]
    \centering
    \includegraphics[width=0.8\textwidth, trim={1.20cm 4.8cm 2.0cm 8.0cm}, clip]{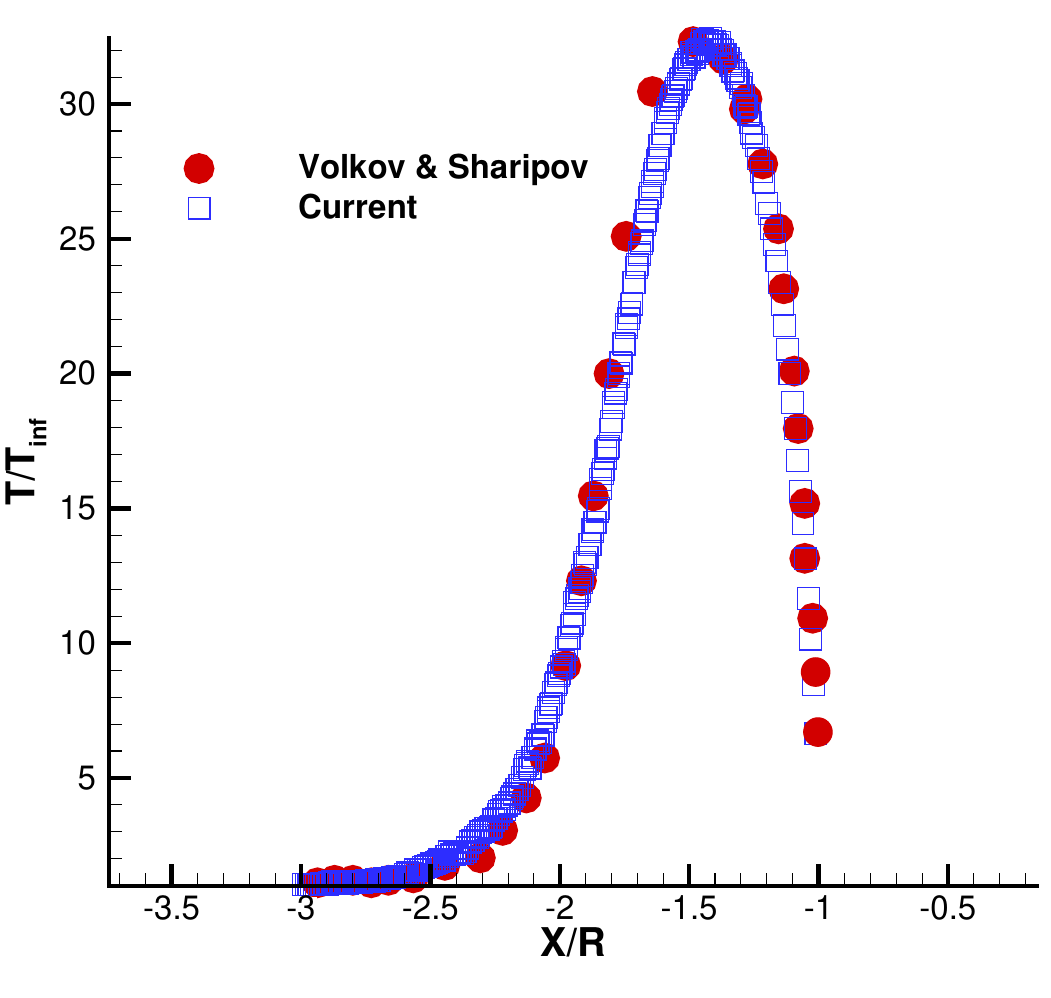}
    \caption{Validation of surface aerodynamic properties against the benchmark results of Volkov and Sharipov \cite{volkov2012direct} for hypersonic Argon flow over a cylinder.}
    \label{fig:sharipov_validation}
\end{figure}

Figure \ref{fig:sharipov_validation} presents the comparison of the surface aerodynamic coefficients (Skin Friction, $C_f$, and Pressure Coefficient, $C_p$) along the cylinder surface. As observed in Figure \ref{fig:sharipov_validation}, the results obtained from our solver show excellent agreement with the benchmark data. The solver accurately captures the peak pressure at the stagnation point ($\theta = 180^\circ$) and correctly resolves the shear stress evolution from the stagnation point to the flow separation region. This validation confirms that the proposed numerical framework correctly resolves the momentum and energy transfer at the gas-surface interface in the transitional hypersonic regime.

\subsection{Machine learning ab-initio vs exact ab-initio}
\begin{figure}[h!]
    \centering
    \includegraphics[width=0.8\textwidth]{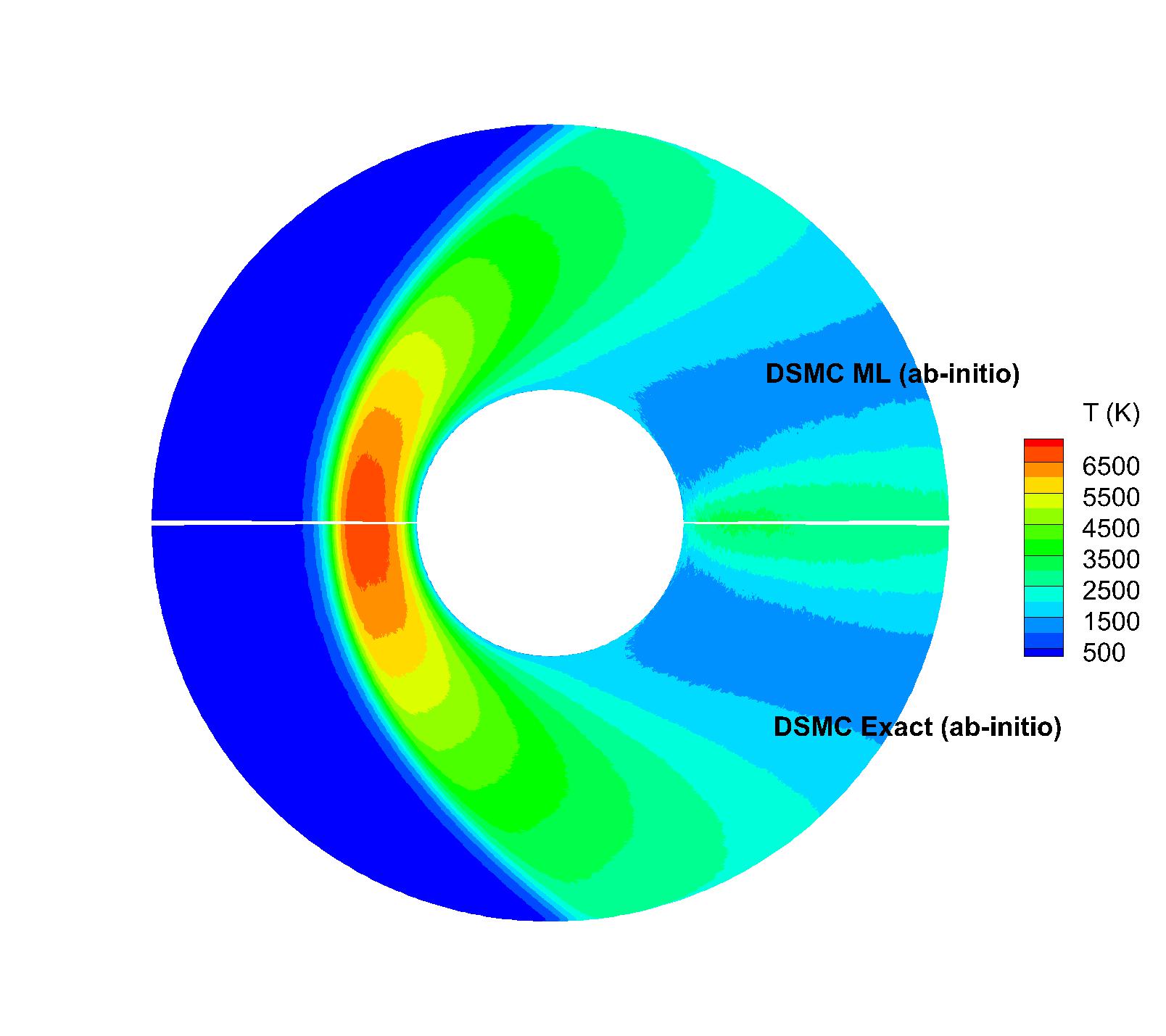}
    \caption{Temperature field comparison for Mach 10 Argon flow. Upper half: Neural Network Surrogate; Lower half: Exact Numerical Integration. The continuity across the centerline demonstrates that the NN accurately captures the shock standoff distance and the post-shock thermal relaxation, reaching a peak temperature of approx $6800$ K.}
    \label{fig:temp_contour}
\end{figure}

\begin{figure}[h!]
    \centering
    \includegraphics[width=0.8\textwidth]{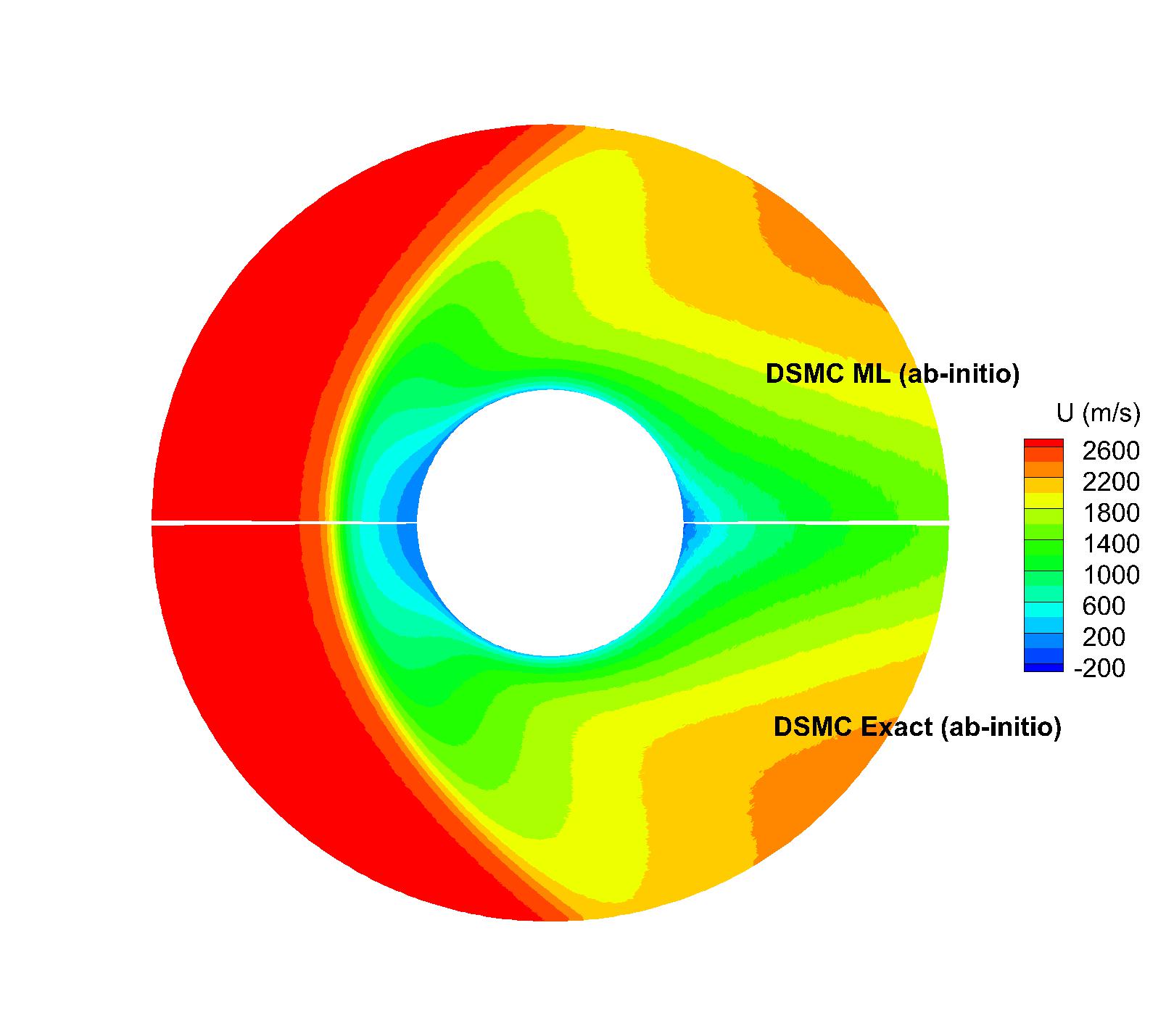}
    \caption{Velocity magnitude ($U$) contours. Upper half: Neural Network Surrogate; Lower half: Exact Numerical Integration. The velocity stagnation point at the cylinder nose and the wake region structure are reproduced identically by the ML model.}
    \label{fig:vel_contour}
\end{figure}

\begin{figure}[h!]
    \centering
    \includegraphics[width=0.8\textwidth]{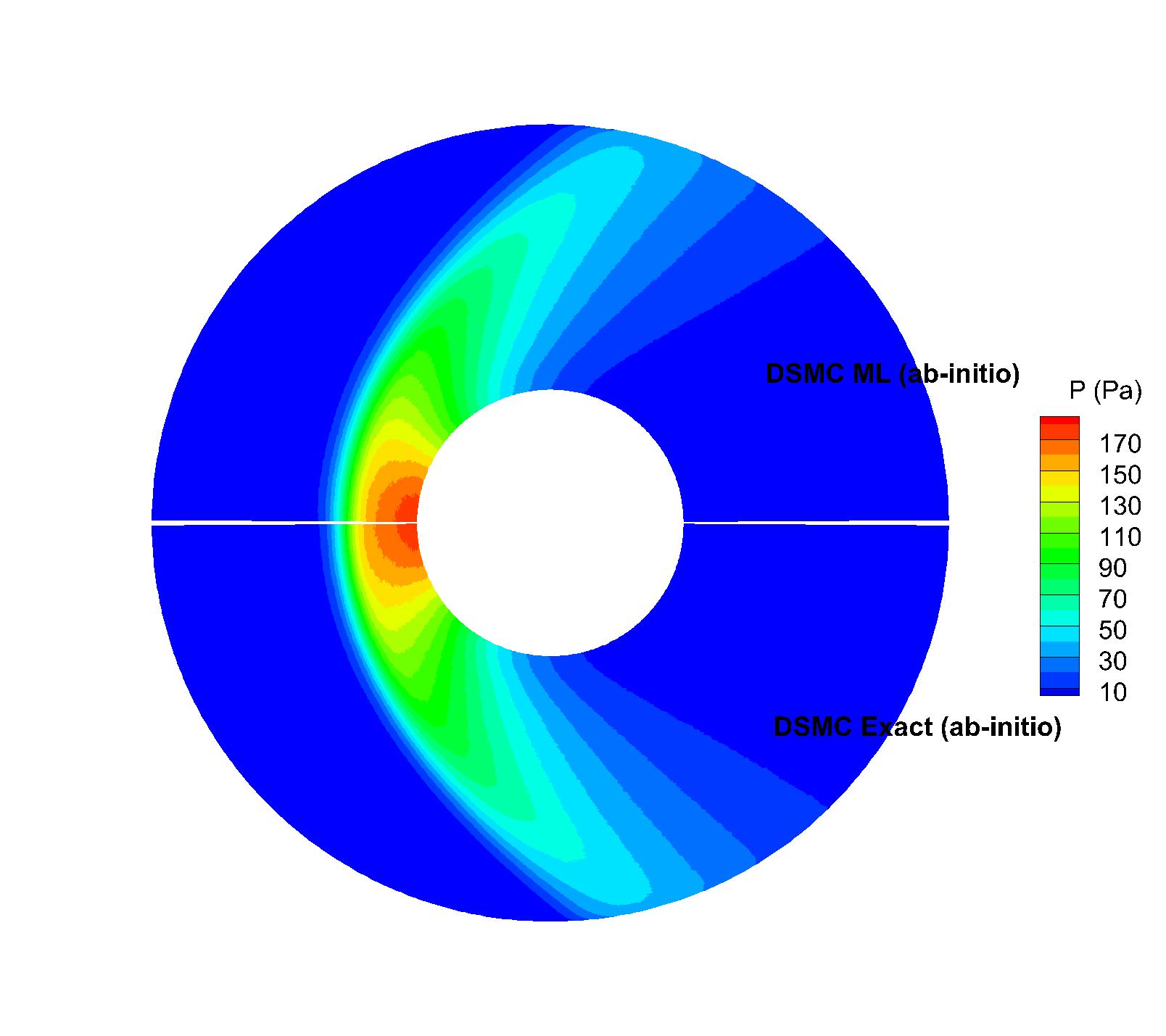}
    \caption{Pressure field comparison. Upper half: Neural Network Surrogate; Lower half: Exact Numerical Integration. The sharp pressure gradient across the shock wave indicates that the neural operator introduces negligible numerical diffusion.}
    \label{fig:press_contour}
\end{figure}

Figures \ref{fig:temp_contour}, \ref{fig:vel_contour}, and \ref{fig:press_contour} present a direct visual comparison of the macroscopic flow fields for cylinder flow at Kn=0.01 from "machine learning ab-initio" vs "exact ab-initio". In each figure, the upper half represents the Neural Network solution, while the lower half shows the Exact benchmark.

Fig. \ref{fig:temp_contour} shows that the bow shock structure is captured with remarkable precision. The high-temperature region behind the shock, where temperatures exceed $6500$ K, is identical in both methods. This confirms that the neural network correctly learned the high-energy repulsive scattering (verified in the $T>5000$ K validation plots), ensuring accurate conversion of kinetic energy into thermal energy without artificial numerical heating.

The velocity contours shown in Fig.~\ref{fig:vel_contour}) show an excellent agreement in the stagnation region and the acceleration zone over the cylinder shoulder. The seamless transition across the symmetry line proves that the momentum transfer cross-sections predicted by the neural operator are physically consistent with the exact \textit{ab initio} potential.

The pressure contour flood, shown in Fig. \ref{fig:press_contour}, and which is sensitive to the correct collision rate and scattering angle distribution, shows no visible discrepancy. The shock thickness and location are indistinguishable between the two methods, validating the "Physics Harvesting" training strategy.

\subsubsection{Surface Properties Verification}
To quantify the accuracy at the gas-surface interface, we compared the skin friction coefficient ($C_f$) distribution along the cylinder surface.

\begin{figure}[h!]
    \centering
    \includegraphics[width=0.8\textwidth]{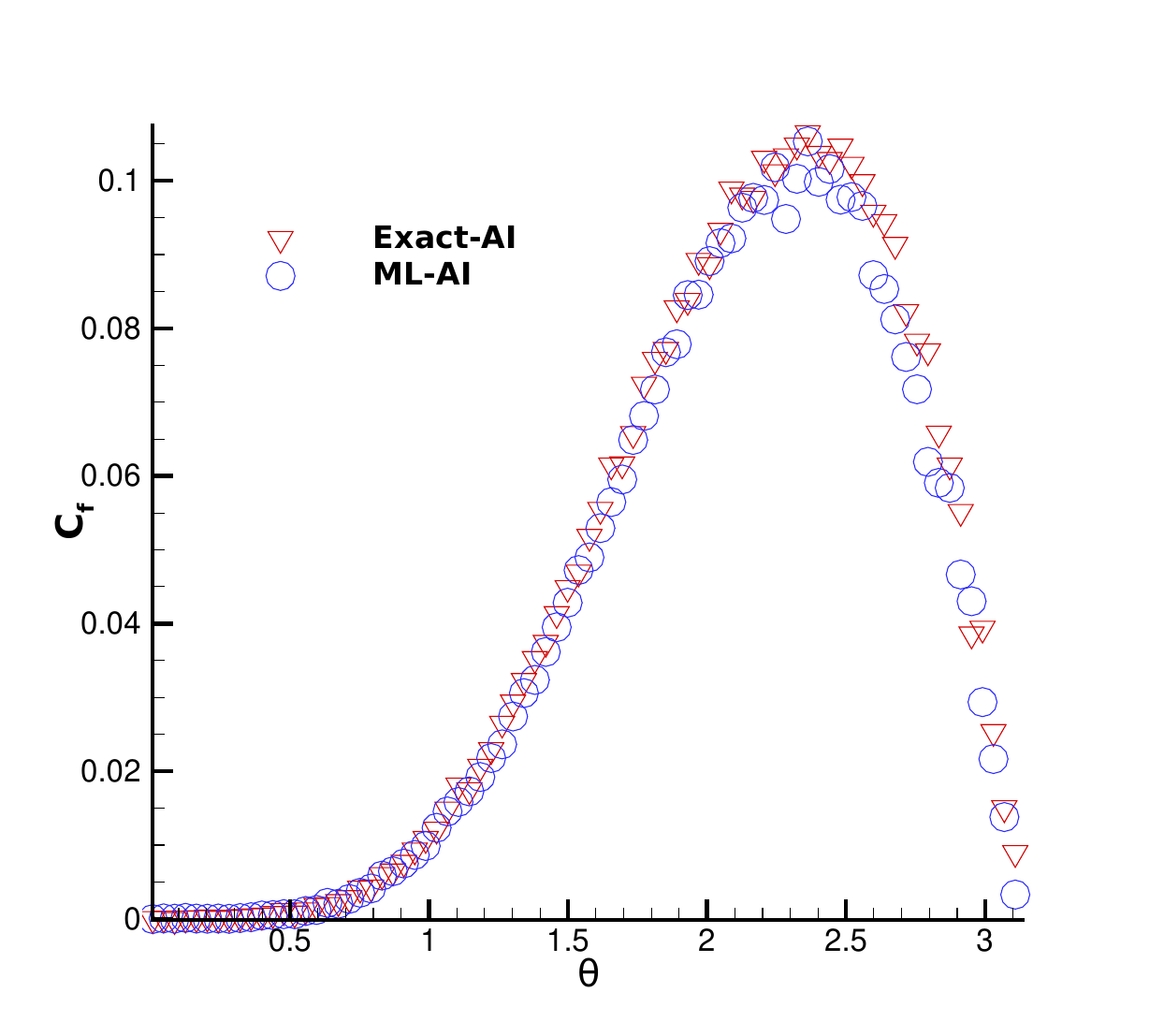}
    \caption{Distribution of skin friction coefficient ($C_f$) along the cylinder surface ($\theta$). Red triangles: Neural Network Surrogate; Blue circles: Exact Numerical Integration. The close agreement confirms that the viscosity and velocity slip at the wall are correctly resolved by the neural operator.}
    \label{fig:cf_comparison}
\end{figure}

As shown in Figure \ref{fig:cf_comparison}, the $C_f$ profile predicted by the ML solver closely follows the Exact solution from the stagnation point ($\theta=0$) to the separation region in the wake. The peak shear stress location and magnitude are accurately captured. Minor statistical fluctuations are observed, which are inherent to the stochastic nature of DSMC sampling, but no systematic bias is evident. This confirms that the neural operator correctly reproduces the transport properties (viscosity) even in the near-wall Knudsen layer where non-equilibrium effects are dominant.

\subsubsection{Computational time efficiency}

The simulation of the exact ab-initio case required around 8 hours of GPU on a NVIDIA A100-SXM4-80GB. The most time-consuming part was to create table of data for ab-initio use during the simulations. However, the simulation of the same test case took around 6:15 hours when the machine learning trained ab-initio framework was employed. Using the machine-learning-trained ab-initio framework reduced the runtime from 8.0 hours to 6.25 hours, corresponding to a 21.9 percent time improvement (speedup). 

The reported runtime reduction (from 8.0~h to 6.25~h, i.e., 21.9\%) corresponds to the total wall-clock time of the full DSMC simulation, which includes particle advection, cell sorting, boundary interactions, and other components that are identical in both solvers.

Interpreting this number must therefore follow Amdahl's law: if only a fraction $p$ of the total runtime is attributable to the collision/scattering subroutine, then even a large acceleration of that subroutine yields a bounded improvement in end-to-end wall time,
\begin{equation}
S_{\text{total}}=\frac{1}{(1-p)+\dfrac{p}{S_{\text{coll}}}} \, .
\end{equation}

Based on our code profiling, the exact scattering integral evaluation accounts for $p \approx 25\%$--$30\%$ of the total unoptimized runtime. Since the neural table look-up reduces this specific kernel cost to near-zero, the theoretical maximum end-to-end speedup is bounded by
\begin{equation}
S_{\max} \le \frac{1}{1-p},
\end{equation}
which yields
\begin{equation}
S_{\max} \approx \frac{1}{1-0.25} = 1.33
\qquad \text{to} \qquad
S_{\max} \approx \frac{1}{1-0.30} = 1.42.
\end{equation}
Equivalently, this corresponds to a maximum achievable total runtime reduction of $p$ (i.e., $25\%$--$30\%$) when the remaining portions of the code are unchanged. Our observed end-to-end time reduction of $21.9\%$ is therefore strongly and realistically consistent with this theoretical bound, once practical factors such as memory transfer and integration overheads are taken into account.

In our case, the neural \textit{ab initio} framework targets precisely the portion of the code that is expensive and physics-specific (scattering evaluation and table construction), while the remaining DSMC pipeline remains unchanged by design. Consequently, the 21.9\% gain should be viewed as an end-to-end improvement under a realistic workflow rather than a micro-benchmark of a single kernel.

For completeness and transparency, we also emphasize that the speedup of the collision/scattering component alone is substantially higher than the end-to-end gain, because the surrogate removes repeated integral evaluations and replaces them with constant-time GPU table queries. Reporting both (i) end-to-end wall-clock speedup and (ii) collision-kernel-only speedup provides a clear and fair assessment of performance improvements in a production DSMC solver.

\section{Conclusions}
This work presented a physics-constrained neural-operator framework that accelerates Direct Simulation Monte Carlo (DSMC) simulations while maintaining the stochastic behavior and conservation consistency required for long-time kinetic computations. Rather than replacing DSMC end-to-end, the proposed approach targets the dominant cost and modeling bottleneck---the collision scattering rule---and substitutes it with learned operators that are constrained by physical structure.

For engineering-level collision modeling, we introduced a cell-local neural collision surrogate that replaces the Variable Hard Sphere (VHS) scattering mechanism inside an otherwise standard DSMC loop. The surrogate is formulated as a compact pairwise MLP evaluated on local collision candidates within DSMC cells, enabling scalable execution without the quadratic cost of global interaction modeling. A central finding is that naive deterministic inference is not thermodynamically stable: repeated regression-style updates suppress velocity variance, producing an unphysical cooling instability and a collapse of the velocity distribution. To eliminate this pathology, we developed a physics-constrained stochastic inference layer that (i) reintroduces controlled fluctuation content through latent-noise injection and (ii) restores conservation consistency through a cell-wise moment-matching correction. This correction is lightweight, non-learnable, and compatible with the DSMC collision partition, making it straightforward to integrate into existing solvers. With this augmentation, the learned collision operator remains stable over long horizons and preserves physically meaningful statistics beyond the mean flow.

For high-fidelity physics, we addressed the prohibitive cost of quantum-mechanical scattering by training a dedicated ab initio neural operator for the J\"ager potential using a physics-harvesting data strategy over large collision-pair ensembles. The resulting surrogate provides a fixed-cost approximation to scattering outcomes while retaining the accuracy needed in hypersonic rarefied conditions. Validation on hypersonic argon flow over a cylinder at Mach~10 demonstrated that the combined framework reproduces key shock features and transport behavior with high fidelity while delivering substantial computational savings relative to direct numerical evaluation of the scattering integral.

Overall, the main contributions are threefold: (1) a practical, drop-in local neural collision surrogate for DSMC collision updates; (2) a physics-constrained stochastic stabilization mechanism that prevents regression-to-the-mean cooling and enforces conservation consistency within DSMC cells; and (3) a scalable ab initio neural operator that substantially reduces the cost of high-energy scattering physics in hypersonic regimes. These results support the broader conclusion that physics-constrained neural operators can accelerate DSMC across both phenomenological and ab initio collision modeling settings, enabling stable, high-throughput rarefied-flow simulations for complex geometries and extreme flight conditions.

\section*{ACKNOWLEDGMENTS}
Stefan Stefanov was partially supported by the MES under the Grant No. D01-325/01.12.2023 for NCDSC – part of the Bulgarian National Roadmap on RIs.

\section*{Declaration of Interests} The authors report no conflict of interest.

\bibliographystyle{unsrtnat}

\bibliography{references} 

\end{document}